\documentclass[pre,twocolumn,preprintnumbers,amsmath,amssymb,showpacs,nofootinbib,floatfix]{revtex4}

\usepackage{graphicx,bm}

\makeatletter
\def\graphicscale{\twocolumn@sw{0.3}{0.4}}
\def\graphicthreescale{\twocolumn@sw{0.3}{0.4}}

\begin{document}

\title{Dynamic finite-size scaling at first-order transitions}

\author{Andrea Pelissetto}
\affiliation{Dipartimento di Fisica dell'Universit\`a di Roma ``La Sapienza"
        and INFN, Sezione di Roma I, I-00185 Roma, Italy}

\author{Ettore Vicari}
\affiliation{Dipartimento di Fisica dell'Universit\`a di Pisa
        and INFN, Sezione di Pisa, I-56127 Pisa, Italy}

\date{\today}

\begin{abstract}
We investigate the dynamic behavior of finite-size systems close to a
first-order transition (FOT).  We develop a dynamic finite-size
scaling (DFSS) theory for the dynamic behavior in the coexistence
region where different phases coexist. This is characterized by an
exponentially large time scale related to the tunneling between the
two phases. We show that, when considering time scales of the order of
the tunneling time, the dynamic behavior can be described by a
two-state coarse-grained dynamics.  This allows us to obtain exact
predictions for the dynamical scaling functions. To test the general
DFSS theory at FOTs, we consider the two-dimensional Ising model in
the low-temperature phase, where the external magnetic field drives a
FOT, and the 20-state Potts model, which undergoes a thermal FOT.
Numerical results for a purely relaxational dynamics fully confirm the
general theory.

\end{abstract}

%\pacs{05.70.Fh,05.70.Ln,64.60.Ht,05.50.+q}

\maketitle

%64.60.Ht 	Dynamic critical phenomena
%05.30.Rt       Quantum phase transitions
%05.70.Fh       phase transitions 
%05.70.Jk       Critical phenomena    
%64.60.-i       general studies of phase transitions
%64.60.an       Finite size systems                                            
%64.70.qj 	Dynamics and criticality
%64.60.an finite size systems
%64.60.De Statistical mechanics of model systems 
%64.60.Ht Dynamic critical phenomena
%05.50.+q 	Lattice theory and statistics (Ising, Potts, etc.) 
%05.70.Ln 	Nonequilibrium and irreversible thermodynamics 
%05.30.Rt       Quantum phase transitions
%05.70.Fh       phase transitions 
%05.70.Jk       Critical phenomena    
%64.60.-i       general studies of phase transitions
%64.60.an       Finite size systems                                            
%64.70.qj 	Dynamics and criticality
%64.60.an finite size systems
%64.60.De Statistical mechanics of model systems 
%64.60.Ht Dynamic critical phenomena

% ========================= BODY =========================

\section{Introduction}

Close to a phase transition point finite-size systems exhibit a
universal finite-size scaling (FSS)
behavior~\cite{FB-72,Barber-83,Cardy-88,Privman-90,PHA-91,PV-02,CPV-14},
which characterizes both static and dynamic equilibrium properties.
FSS is also observed in out-of-equilibrium phenomena, for instance, in
the quenching of a random configuration at the critical point. In
general, it is observed when the size $L$ of the system is larger than
any microscopic length scale and if the observation time $t$ is
comparable with the time scale $\tau(L)$ of the slowest critical mode,
which generally diverges in the infinite-volume limit~\cite{HH-77}.
At continuous transitions the finite-size behavior is characterized by
power laws, with universal critical exponents which only depend on a
few global features of the system; see, e.g.,
Refs.~\cite{CG-04,GZHF-10}.  A static FSS is also observed at
first-order transitions
(FOTs)~\cite{NN-75,FB-82,PF-83,FP-85,CLB-86,Binder-87,BK-90,VRSB-93,%%
  CCPV-04,CNPV-14}.  In this case one observes power-law behaviors
with simple exponents, which are closely related to the space
dimension of the system.

At FOTs, dynamic phenomena play a very important role, due to the
presence of very slow modes with large time scales.  Indeed, in the
absence of continuous symmetries, any local dynamics is very slow, due
to an exponentially large tunneling time between the two phases
coexisting at the transition point: $\tau(L) \sim \exp(\sigma
L^{d-1})$ for a system of size $L^d$, where the constant $\sigma$ is
generally related to the interface free energy.  Also the dynamic
behavior, both in equilibrium and out-of-equilibrium conditions, is
supposed to show universal features and, in particular, to exhibit a
universal dynamic FSS (DFSS). A satisfactory understanding of the DFSS
properties of the system close to a FOT is important for experiments
on relatively small systems, when the longest time scale of the system
is of the order of the time scale of the experiment.

In this paper we consider the evolution of a finite-size system close
to a FOT.  We focus on the interplay between the finite size of the
system and the distance (in parameter space) from the FOT point.  We
show that several large-scale quantities obey DFSS laws, analogous to
those holding at continuous transitions, the only difference being
that the time scale $\tau(L)$ increases exponentially with $L$.
Moreover, as long as the control parameters (for instance,
temperature, magnetic field, $\ldots$) are such the system is always
in the coexistence region, the observed behavior can be interpreted in
terms of a generic Markov two-state coarse-grained dynamics.  Using
such dynamics, we can derive exact predictions for the DFSS functions.
To test the general DFSS theory, we present numerical analyses of the
two-dimensional (2D) Ising model in the low-temperature phase---here
the external magnetic field drives a FOT---and of the 2D 20-state
Potts model, which undergoes a thermal FOT.  Some related issues are
investigated in Refs.~\cite{PV-17,LZ-17}, where the off-equilibrium
behavior observed when some parameter is slowly varied across a FOT
(the analogue of the Kibble-Zurek dynamics at a continuous
transition~\cite{Kibble-76, Zurek-85}) is investigated.

The paper is organized as follows. In Sec.~\ref{sec2} we consider the
two-dimensional Ising model, define the relevant observables, and the
dynamics that we consider. The general scaling theory is developed in
Sec.~\ref{sec3} and tested in Sec.~\ref{sec4}. In Sec.~\ref{sec5} we
discuss a different type of finite-size scaling that allows us to
investigate the single-droplet region. In Sec.~\ref{sec6} we extend
the general discussion to the case in which the magnetic field does
not vanish only on a subset of lattice points.  In
Sections~\ref{sec7}, \ref{sec8}, \ref{sec9}, \ref{sec10} we extend the
discussion to the Potts model. In Sec.~\ref{sec11} we summarize and
draw our conclusions.  In the Appendix \ref{AppA} we report the
computation of the average magnetization for an Ising model in which
the magnetic field is nonvanishing only in a single site. In
App.~\ref{AppB} we compute the average energy for the Potts model at
the transition in the presence of a single strongly ferromagnetic
bond.

\section{The Ising case: Definitions} \label{sec2}

\subsection{The model} \label{sec2.A}

We consider the 2D Ising model defined on a square $L\times L$ lattice
in the presence of an external magnetic field. Its Hamiltonian is
\begin{equation}
H = - \sum_{\langle ij\rangle} s_i s_j - h \sum_i s_i, 
\label{isiham}
\end{equation}
where $s_i=-1,\,1$ and the first sum is over all nearest-neighbor
pairs $i,j$.  The model undergoes a paramagnetic-ferromagnetic
transition for $h=0$ and $T=T_c$, with~\cite{McCoy-book}
\begin{equation}
\beta_c = {1\over 2} \ln(1 + \sqrt{2}),\qquad T_c = 1/\beta_c.
\label{crittemp}
\end{equation}
For $T<T_c$ and $h\to 0$ the system is spontaneously magnetized in the
thermodynamic limit. The spontaneous magnetization per site is given
by
\begin{equation}
m_0(T) = \left[ 1  - \sinh(2\beta)^{-4}\right]^{1/8}.
\label{magne}
\end{equation}
In the following we also need the interface tension $\kappa$,
which is also  known exactly~\cite{ZA-82}:
\begin{equation}
\kappa = 2 + \ln[\tanh(\beta)]/\beta. 
\label{intten}
\end{equation}
In a finite square box of linear size $L$, the behavior of the system
depends on the boundary conditions. For boundary conditions that
preserve the $\mathbb{Z}_2$ inversion symmetry, for instance, for
periodic boundary conditions (PBC), the magnetization vanishes for
$h=0$. For small values of $h$, static FSS holds in terms of the
scaling variable
\begin{equation}
r_1 = h L^2.
\label{r1def}
\end{equation}
This means that an appropriate universal behavior is observed when taking 
the limits $h\to 0$, $L\to\infty$ at fixed $r_1$. In particular, 
in the FSS limit the 
magnetization per site $m$ becomes 
\begin{equation}
m = m_0 \, f_{\rm eq}(r_1),\qquad
f_{\rm eq}(r_1) = {\rm tanh} (\beta m_0 r_1 ).
\label{feqs} 
\end{equation}
Note that $m \not= |m_0|$ for any finite $r_1$,
indicating that both free-energy minima contribute to equilibrium
properties, i.e., that the system is always in the coexistence region.

\subsection{The dynamics} \label{sec2.B}

We consider a purely relaxational dynamics at fixed $T < T_c$ and
fixed magnetic field $h$.  We use three different implementations of
the Metropolis algorithm, which, as we shall see, all show the same
dynamical behavior. In most of the simulations we use the checkerboard
update. If $(n_x,n_y)$, $0\le n_x,n_y<L$, are the coordinates of the
lattice sites, we first update all spins at points such that $n_x +
n_y$ is even (the order is irrelevant since they do not interact),
then all spins at points such that $n_x + n_y$ is odd. We also
consider a sequential update, in which we first sequentially update
all spins on the line $n_y=0$, then those on the line $n_y = 1$, and
so on. Finally, we consider the random update, in which spins are
randomly chosen. All times are measured in sweeps. In the checkerboard
and sequential updates, a sweep consists in a Metropolis update
attempt of all spins. In the random case, it consists in $L^2$ random
update attempts.  In all cases, we start the dynamics from a
completely ordered configuration with $s_i = -1$ for all $i$'s.

The main purpose of the paper is that of verifying the existence of a
DFSS behavior, which extends the static FSS to the dynamics when $h$
is small and $T < T_c$. As the relevant scaling variable is expected
to be $r_1 = h L^2$, simulations are performed at fixed $T$ and $r_1$
for different values of $L$, varying at the same time the magnetic
field as $h = r_1/L^2$. Note that $h\to 0$ as $L$ increases.

In the evolution we measure the average magnetization per site
\begin{equation}
    M(t) = {1\over L^2} \sum_i s_i,
\end{equation}
where $t$ is the time,  and the corresponding  average
renormalized magnetization
\begin{equation}
     m_r(t,r_1,L) = {1\over m_0} \langle M(t) \rangle,
\end{equation}
where the average is over the different dynamic histories and we have
not reported explicitly the temperature dependence. Moreover, given a
number $\mu$ satisfying $-1 < \mu < 1$, we define the first-passage
time $t_{f}(\mu)$ as the smallest time such that
\begin{equation}
     M[t_{f}(\mu)] = \mu m_0.
\end{equation}
We can then consider its average
\begin{equation}
T_{f}(\mu,r_1,L) =  \langle t_{f}(\mu) \rangle,
\end{equation}
and its probability distribution
\begin{equation}
P(x,r_1,L) = \left\langle 
\delta\left( {t_{f}(\mu)\over T_{f}(\mu,r_1,L)} - x \right)
            \right \rangle.
\end{equation}

\section{The Ising case: Dynamic scaling behavior in the coexistence region} 
\label{sec3}

\subsection{General arguments} \label{sec3.A}

Close to the FOT at $h=0$ and $T<T_c$, physical observables show a
scaling behavior in terms of $h$ and of the size of the system $L$,
which depends in general on the boundary conditions. For PBC, the only
case we consider in this work, static observables show FSS once they
are expressed in terms of $r_1=h L^2$.  To extend FSS to the dynamic
case, it is necessary to identify the appropriate time scale of the
dynamics. As we consider the large-$L$ limit at fixed $r_1$, 
the system is always in the coexistence region.
Therefore, the relevant time scale is the one that controls the large-time 
dynamic behavior for $h=0$.

For symmetric boundary conditions, in the low-temperature phase the
largest autocorrelation times are associated with flips of the
magnetization. This should occur by means of the generation of
configurations characterized by two coexisting phases separated by two
approximately planar interfaces. Their probability is of the order of
$\exp(-\sigma L)$, where
\begin{equation}
\sigma = 2 \beta \kappa,
\label{sigrelka}
\end{equation}
and $\kappa$ is the planar interface tension.  The factor of two is
due to the presence of two interfaces, which are necessarily present
because of the PBC.  The time needed to observe a reversal of the
magnetization is proportional to $\exp (\sigma L)$ with power
corrections~\cite{BH-book,MT-85}. Therefore, we define a time scale
\begin{equation}
\tau(L)  = L^\alpha \exp(\sigma L),
\label{tauL}
\end{equation}
where $\alpha$ is an appropriate exponent.  

Note that Eq.~(\ref{tauL}) assumes that the relevant mechanism for the
generation of the opposite phase is the creation of strip-like domains
parallel to the lattice axes and not the creation of spherical
droplets, as it has already been checked for $h=0$, see
Ref.~\cite{BHN-93}.  This reflects the fact that spherical droplets
are unstable.  Indeed, at $h\simeq 0$ (in the FSS limit $h$ scales as
$L^{-2}$) they tend to shrink due to their curvature, taking a time
$t\sim R^2$, where $R$ is their size~\cite{Bray-94}.  Equivalently,
one can note that a critical droplet has a size $R_c$ of the order of
\cite{RTMS-94} $a/h$, so that $ R_c/L = a L/r_1$. Therefore, at fixed
$r_1$ we find $R_c \gg L$, confirming the irrelevance of the droplets
in the limit we are considering here.

Once we have identified the correct time scale, we can introduce the
scaling variables that parametrize the dynamics.  Beside the static
quantity $r_1$, we define
\begin{equation}
   r_2 = t/\tau(L).
\end{equation}
Then, we expect
\begin{eqnarray}
   && m_r(t,r_1,L) \approx f_m(r_1,r_2), \nonumber \\
   && T_{f}(\mu,r_1,L) \approx \tau(L) f_T(r_1,\mu), \nonumber \\
   && P(x,r_1,L)   \approx f_P(r_1,x) .
\label{dynamic-scaling}
\end{eqnarray}

\subsection{Coarse-grained flip dynamics} \label{sec3.B}

The above scaling relations define several scaling functions.  We now
show that they can be exactly predicted.  Let us consider the dynamics
of a single system.  At $t=0$ the magnetization $M(t)$ is equal to
$-1$. As $t$ increases, $M(t)$ rapidly changes and, after a few
iterations, we observe that $M(t) \approx - m_0$, with fluctuations
that decrease as $L$ increases. Then, suddenly, the magnetization
changes sign. In a very short time interval $\Delta t$, with $\Delta t
\ll \tau(L)$, $M(t)$ increases and $M(t) \approx + m_0$ at the
end. Then, the magnetization remains constant for a long time interval
and then, again, in a very short time interval $\Delta t$, we observe
the reversal of the magnetization, obtaining $M(t) \approx -
m_0$. This flipping process continues as $t$ increases, guaranteeing
that the time average of $m(t)$ converges to the value given in
Eq.~(\ref{feqs}) as the run length goes to $\infty$.

Since on time scales of the order of $\tau(L)$ the reversal of the
sign of the magnetization is essentially instantaneous, we can
consider a simpler coarse-grained dynamics.  First, we assume that
$M(t)$ takes only two values, $\pm m_0$. Second, as we expect the
dynamics restricted within each free-energy minimum to be rapidly
mixing, we can assume that the coarse-grained dynamics is Markovian.
Under these conditions, the dynamics is completely parametrized by the
rates $I_+$ and $I_-$ defined by
\begin{eqnarray}
&& P[M(t)=-m_0 \to M(t+dt) = + m_0] = I_+ dt, \nonumber \\
&& P[M(t)=+m_0 \to M(t+dt) = - m_0] = I_- dt, \qquad
\end{eqnarray}
where $P(\cdot)$ is the probability of the considered transition. 

Consider now $N_{\rm tot}$ different dynamic histories and let
$N_+(t)$ be the number of systems for which $M(t) = + m_0$ at time
$t$. Then, we can write
\begin{equation}
dN_+(t) = - N_+(t) I_- dt + [N_{\rm tot} - N_+(t)] I_+dt.
\end{equation}
If we define $n(t) = N_+(t)/N_{\rm tot}$ we obtain the equation
\begin{equation}
{dn\over dt} = - n I_- + (1 - n) I_+.
\label{eqdiff-n}
\end{equation}
Since $n(t=0)=0$, the solution is 
\begin{equation}
n(t) = {I_+\over \lambda} (1 - e^{-\lambda t}), \qquad 
\lambda = I_+ + I_-.
\end{equation}
Then, since $m_r(t) = 2\, n(t) - 1$, we obtain
\begin{equation}
m_r(t) = {I_+ - I_-\over \lambda} - {2 I_+\over \lambda} e^{-\lambda t}.
\end{equation}
For large $t$ we must recover the equilibrium value (\ref{feqs}), which 
implies 
\begin{equation}
    {I_- \over I_+} = e^{-2\beta m_0 r_1}.
\label{ratioImIp}
\end{equation}

Finally, the rate $I_+$ can be related to the first-passage time.
Indeed, first note that, if the dynamics consists in essentially
instantaneous flips, the quantity $T_{f}(\mu,r_1,L)$ is expected to
become independent of $\mu$ in the scaling limit, i.e., we can simply
write $T_f(\mu,r_1,L) \approx \tau_f(r_1,L)$.  Then, since the
probability that the first flip of the magnetization from $-m_0$ to
$+m_0$ occurs in the time interval $[t,t+dt]$ is $\exp(-I_+ t)I_+ dt$,
in the scaling limit we have
\begin{equation}
    {1\over I_+} = \tau_f(r_1,L)
\label{TIp}
\end{equation}
and
\begin{equation}
    P(x, r_1, L) = e^{-x}.
\end{equation}
Relations (\ref{ratioImIp}) and (\ref{TIp}) allow us to rewrite $m_r(t)$ as 
\begin{eqnarray}
&& m_r(t) \approx  f_{\rm eq}(r_1)   -  [1 + f_{\rm eq}(r_1)]
        e^{-t/T_i}, \label{mr-th} \\
&&T_i = {\tau_{f}(r_1,L) \over    1 + e^{-2\beta m_0 r_1}},
\nonumber 
\end{eqnarray}
where $f_{\rm eq}(r_1)$ is the static FSS function (\ref{feqs}).
In the scaling limit we expect that 
\begin{equation}
\tau_f(r_1,L) = \tau(L) f_T(r_1),
\label{tauflsca}
\end{equation}
see Eq.~(\ref{dynamic-scaling}), and therefore $m_r(t)$ becomes a 
universal function of $r_1$ and $r_2$.

Note that all predictions are independent of the sign of $h$ and also
hold when $h < 0$, i.e., when the magnetic field does not favor the
flip of the magnetization. The symmetry of the model under $h\to -h$
implies $I_+(-h) = I_-(h)$, and therefore the relation
\begin{equation}
     {\tau_{f}(r_1,L)\over \tau_{f}(-r_1,L)} = {g_\tau(r_1)\over g_\tau(-r_1)}=
e^{-2 \beta m_0 r_1}.
\label{ratio_r1_rm1}
\end{equation}

We finally mention that field-theoretical renormalization-group
studies of the purely relaxational dynamics in the critical region
below the critical point, thus in the limit $T\to T_c^-$, are reported
in Refs.~\cite{KDS-96,KD-98}.

\section{The Ising case: Monte Carlo results in the coexistence 
region} \label{sec4}

To verify the previous predictions, we perform Monte Carlo simulations
for $T = 0.9T_c$ and several values of $r_1$, ranging from $-1$ to 50.
All data reported in this section are obtained by using the
checkerboard update, except in the last subsection, where we compare
the results for three different updates.

\subsection{Testing the coarse-grained flip dynamics} \label{sec4.A}

\begin{figure}[tbp]
\includegraphics*[scale=\graphicscale,angle=0]{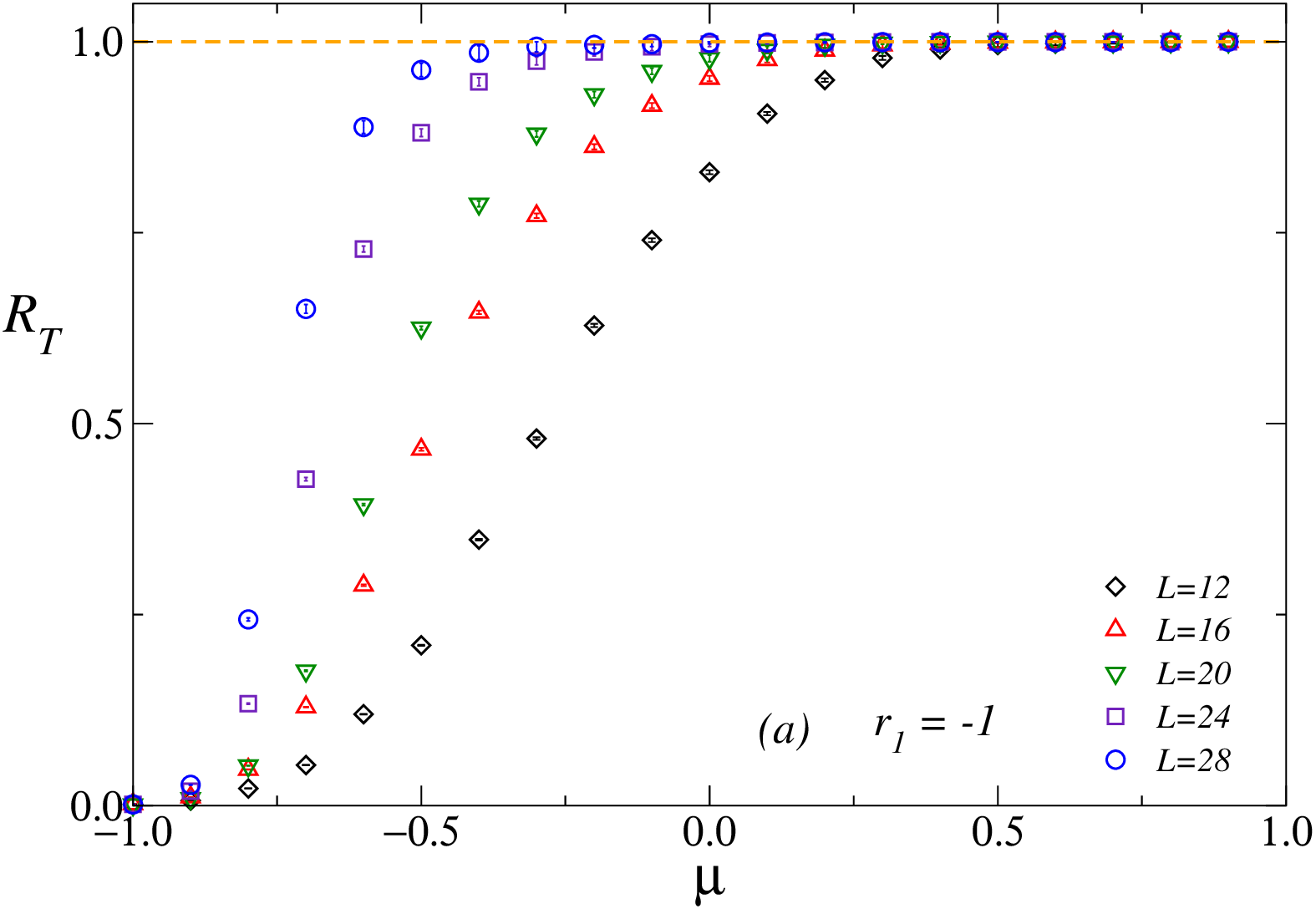}
\includegraphics*[scale=\graphicscale,angle=0]{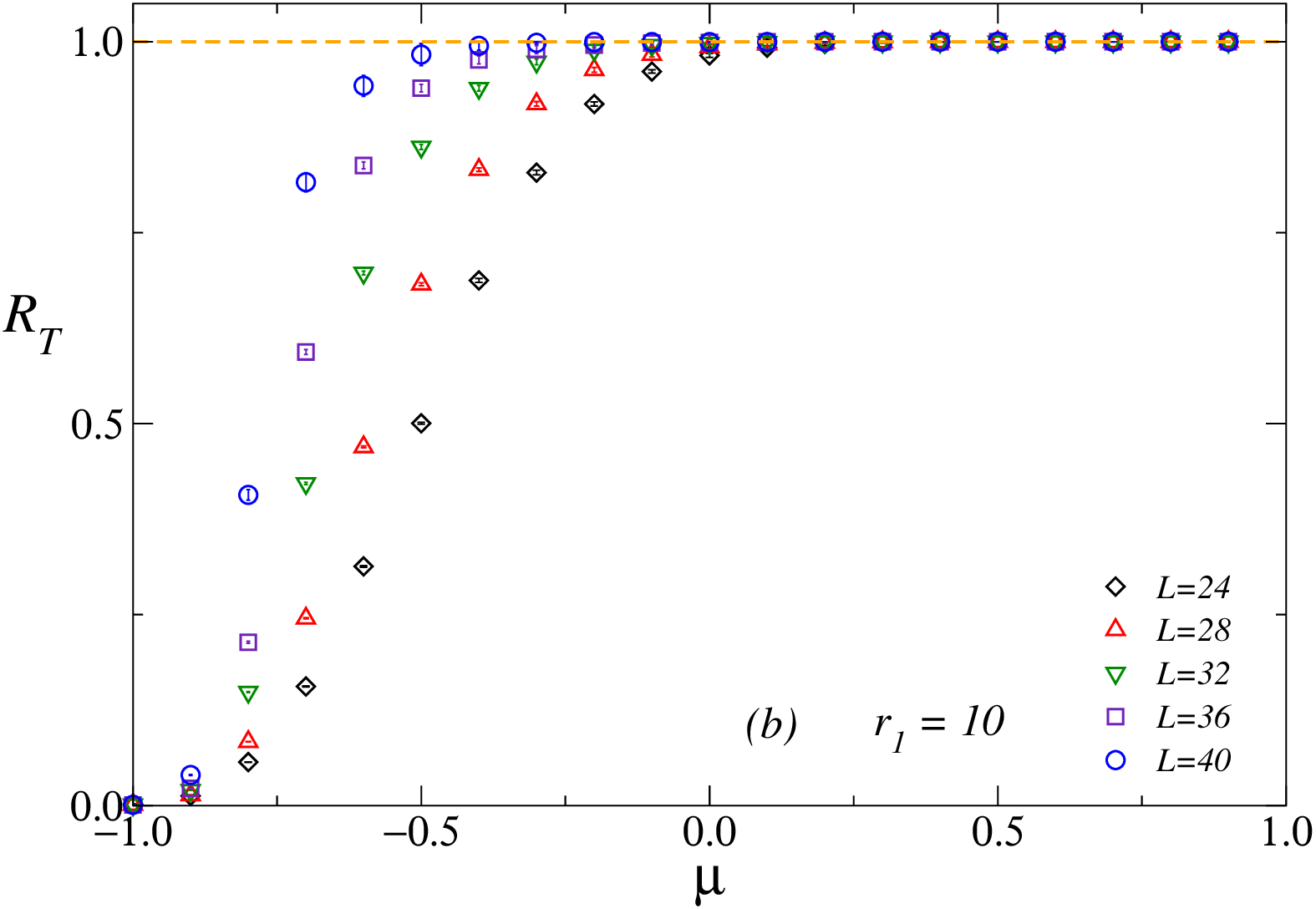}
\includegraphics*[scale=\graphicscale,angle=0]{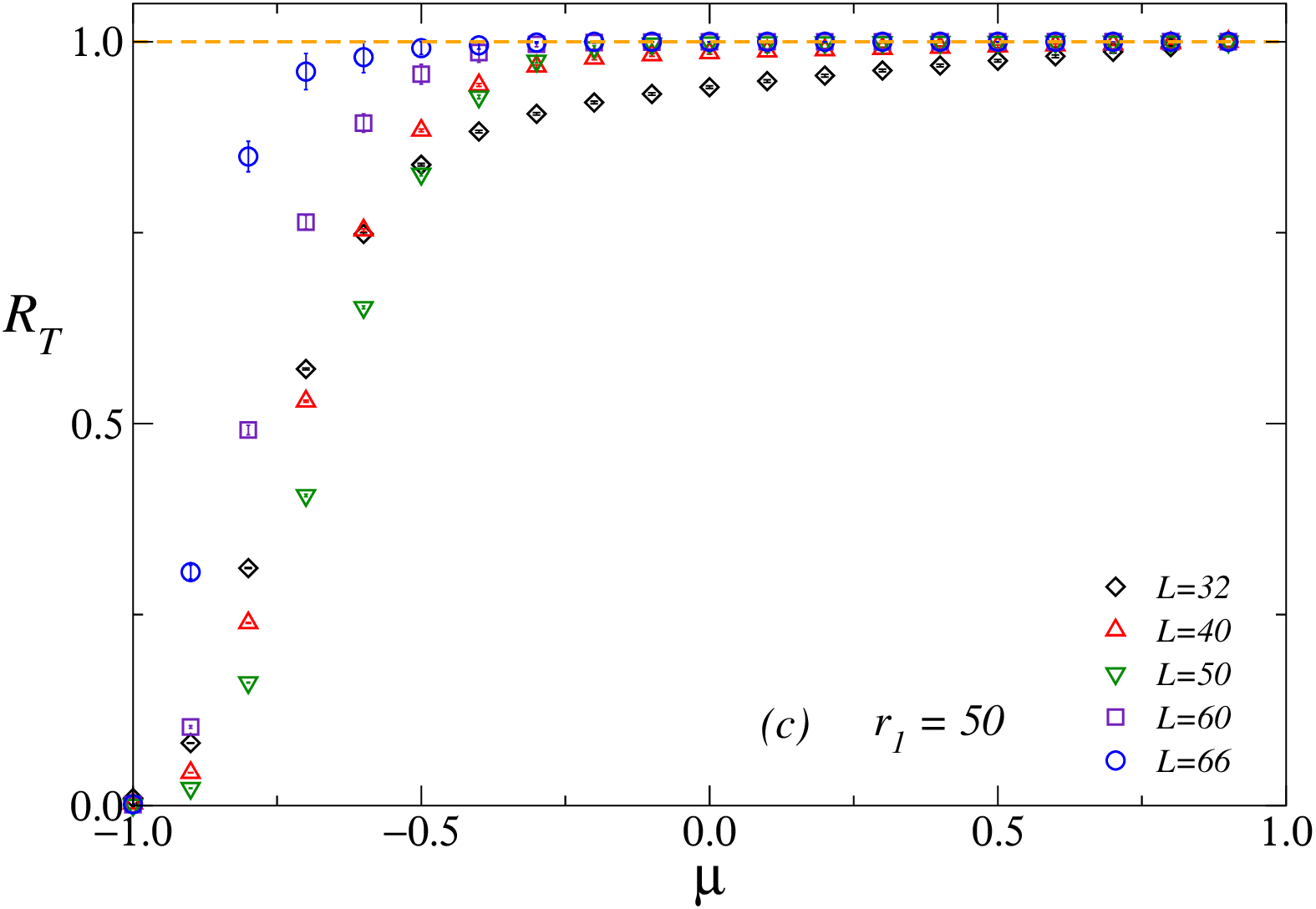}
%=% \vskip-5mm
\caption{Ratio $R_T(\mu,r_1,L)$ for $r_1 = -1,10,50$ and 
several values of $L$, as a function of $\mu$. Here $T = 0.9 T_c$.
}
\label{RT-ratio}
\end{figure}

As a first test, we verify that $T_{f}(\mu,r_1,L)$ becomes independent
of $\mu$ for $L\to \infty$, cf. Eq.~(\ref{TIp}). We consider the ratio
\begin{equation}
R_T(\mu,r_1,L) = {T_{f}(\mu,r_1,L) \over T_{f}(0.9,r_1,L)}.
\end{equation}
Such a quantity is plotted in Fig.~\ref{RT-ratio} for $r_1 =
-1,10,50$.  There are clearly two regimes. For $\mu$ negative and
close to $-1$, the ratio is small. For these values of $\mu$,
$T_{f}(\mu,r_1,L)$ simply gives the typical timescale of the
fluctuations of the magnetization within the free-energy minimum with
$m_r \approx -1$, which is the stable one for $r_1 = - 1$, and the
metastable one for the other two values of $r_1$.  Then,
$T_{f}(\mu,r_1,L)$ becomes essentially constant, which indicates that
these values of the magnetization are only reached in the very rapid
process in which the magnetization changes sign. As $L$ increases,
$R_T(\mu,r_1,L)$ starts to be 1 at decreasing values of $\mu$, a
consequence of the decrease of the fluctuations of the average
magnetization with the volume. For $L\to \infty$ it is then natural to
expect $R_T(\mu,r_1,L)=1$ for any $\mu > -1$.  It is interesting to
observe that the size corrections increase significantly with
$r_1$. For $r_1 = -1$, the ratio at $\mu = 0$ is essentially 1 for
$L\gtrsim 24$, while one should take $L\gtrsim 40$ for $r_1 = 50$.

The numerical data provide also information on the nature of the size
corrections. For this purpose we fit $[1 - R_T(\mu,r_1,L)]$ at fixed
$\mu$ and $r_1$ to $a L^{-p}$.  If we use the data for $r_1 = 10$ ($24
\le L \le 40$) we obtain $p = 5.3(2)$, 5.9(5), 6.9(1.0), 7.3(1.3) for
$\mu = -0.3,-0.2,-0.1,0$, respectively. Similarly large powers are
obtained if one considers other values of $r_1$. The very large values
obtained for $p$ make a power behavior rather unlikely. We have also
tried to parametrize the scaling corrections as $a e^{-bL}$. For $r_1
= 10$ we obtain $b = 0.25(5)$, 0.25(3), 0.22(2), 0.20(1), for the same
values of $\mu$ as before. The $\chi^2$ is slightly better then that
obtained in the power-law fit, which makes the exponential convergence
more plausible than the power-law behavior. It is interesting to note
that the prefactor $b$ appears to be independent of $r_1$, within
errors (say, within 10-15\%).  For instance, for $\mu = -0.2$ we
obtain $b = 0.233(5)$ for $r_1 = -1$ ($L\ge 12$) and $b = 0.20(2)$ for
$r_1 = 50$ ($L\ge 40$).

As a second test of the general theory, we verify the relation
(\ref{ratio_r1_rm1}) by comparing the results for $r_1 = 1$ and $r_1 =
-1$.  We consider the quantity
\begin{equation}
 R_2(L) = {T_{f}(0.9,r_1,L)\over T_{f}(0.9,-r_1,L)} e^{2 \beta m_0 r_1}
\label{r2lrel}
\end{equation}
for $r_1 = 1$. We obtain $R_2(L) = 0.998(4)$ ,0.996(4), 1.008(5),
1.016(8), 0.987(14), for $L=12,16,20,24,28$, respectively.  Therefore,
the data confirm the general relation (\ref{r2lrel}). 

\begin{figure}[tbp]
\includegraphics*[scale=\graphicscale,angle=0]{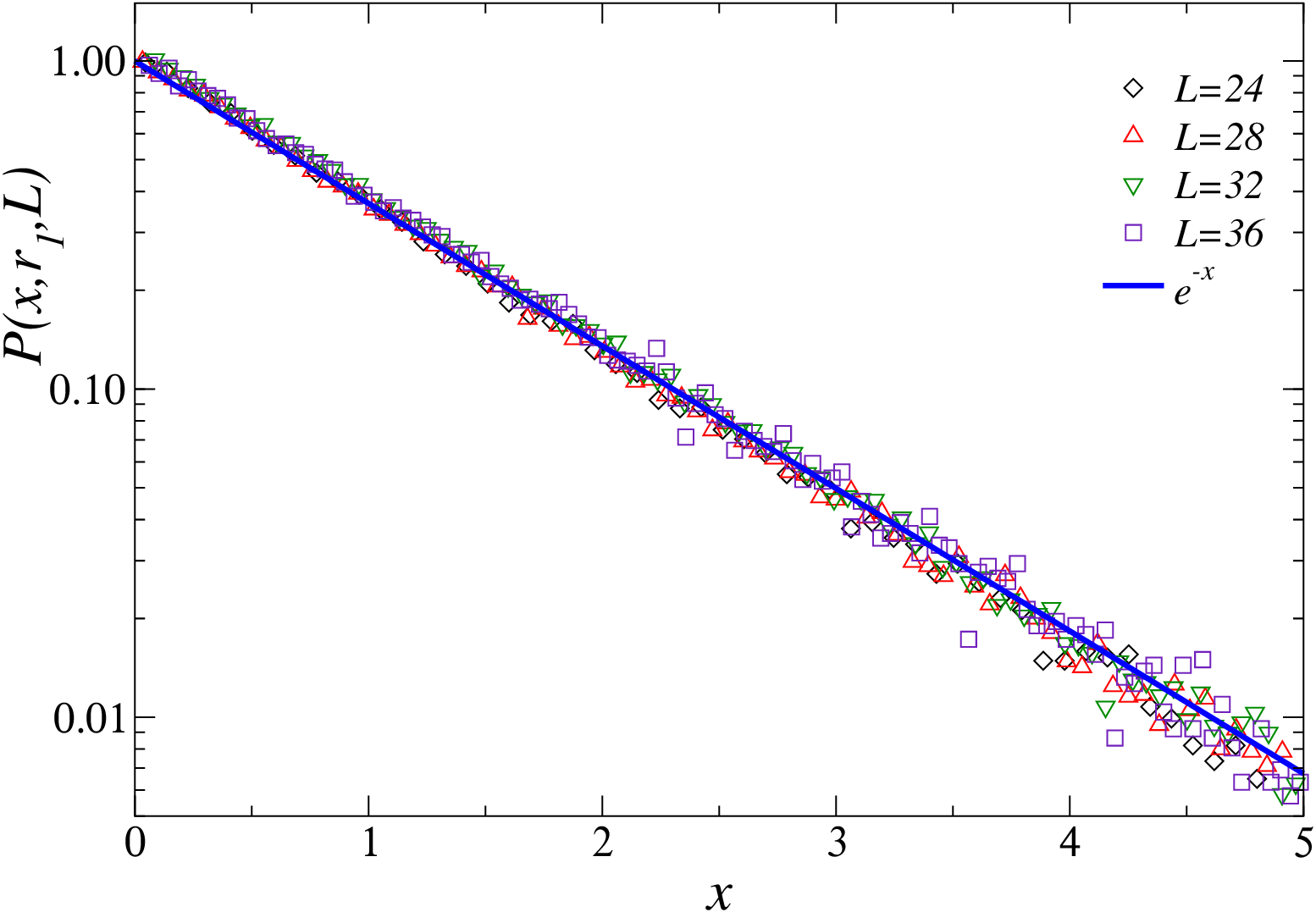}
%=% \vskip-5mm
\caption{Distribution function of the first passage
  time as a function of $x = t_f(0.9)/T_{f}(0.9,r_1,L)$ for $r_1=10$,
  $T = 0.9T_c$, and several values of $L$. The thick blue line 
  corresponds to $e^{-x}$. }
\label{fig:Px}
\end{figure}

We also check the predictions for the distribution function
$P(x,r_1,L)$. Results for $r_1 = 10$ are reported in Fig.~\ref{fig:Px}
together with the theoretical prediction.  Data follow the expected
exponential behavior quite precisely, confirming the two-level nature
of the dynamics.

\begin{figure*}[tbp]
\begin{tabular}{cc}
\includegraphics*[scale=\graphicscale,angle=0]{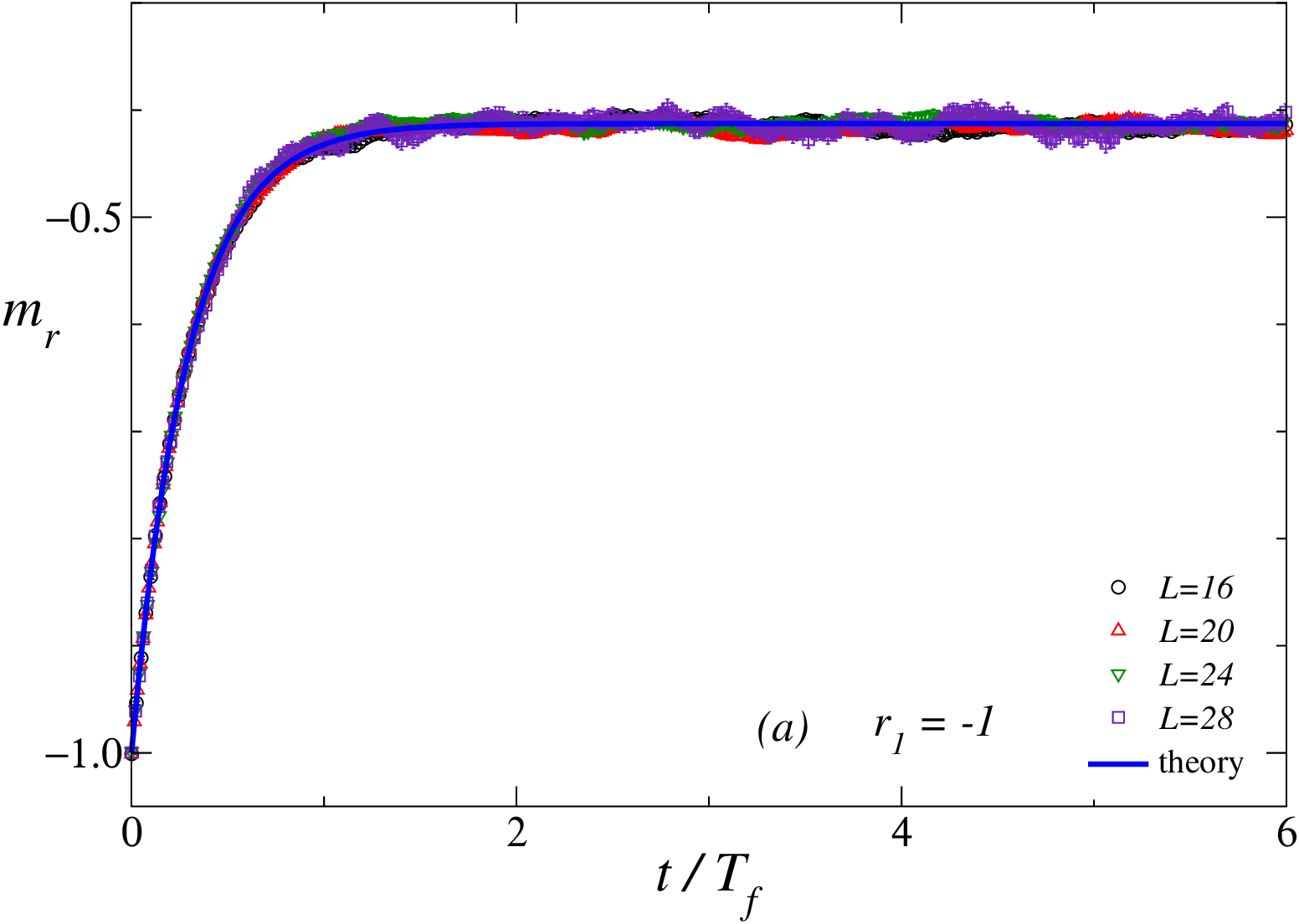} &
\includegraphics*[scale=\graphicscale,angle=0]{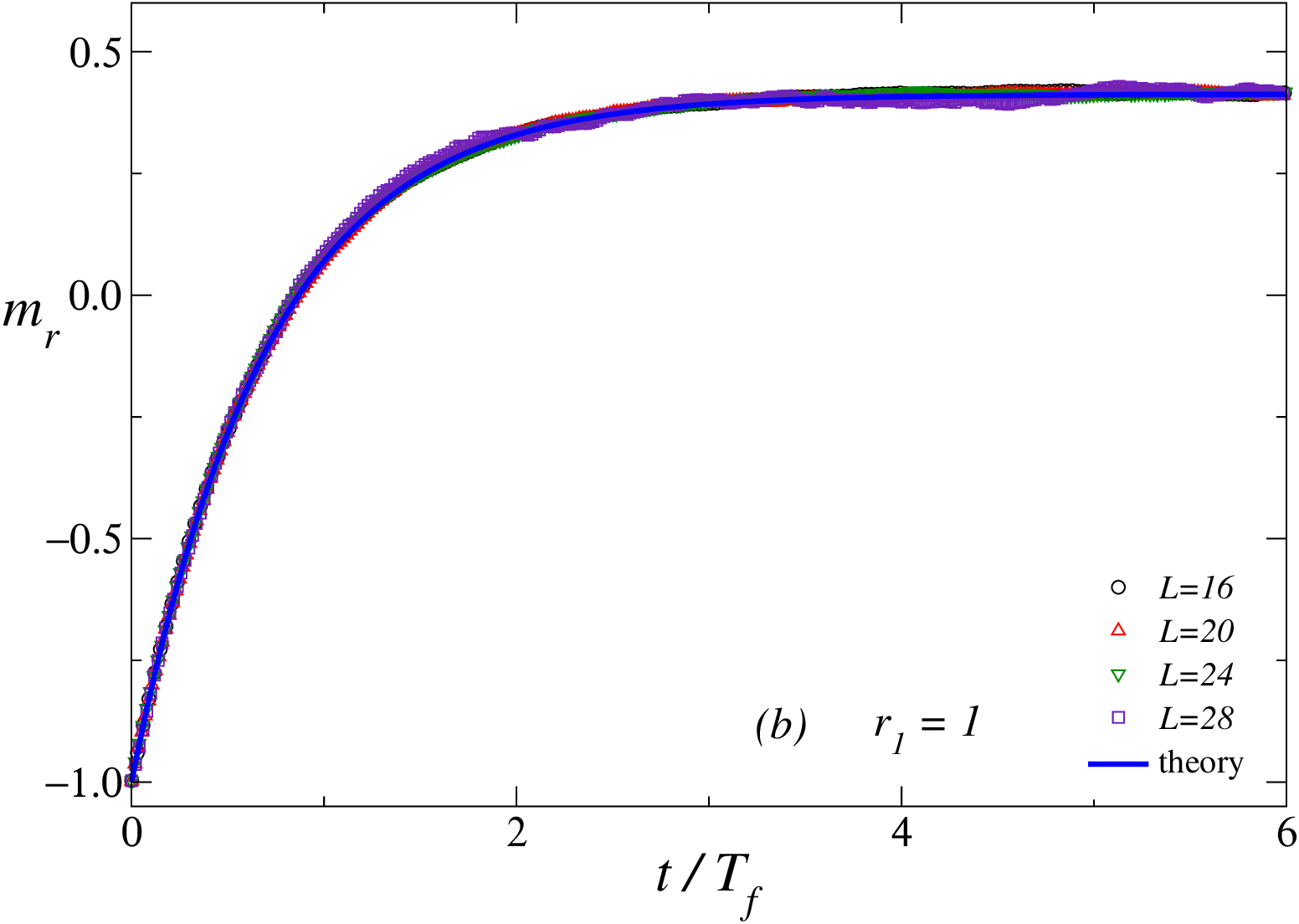} \\
\includegraphics*[scale=\graphicscale,angle=0]{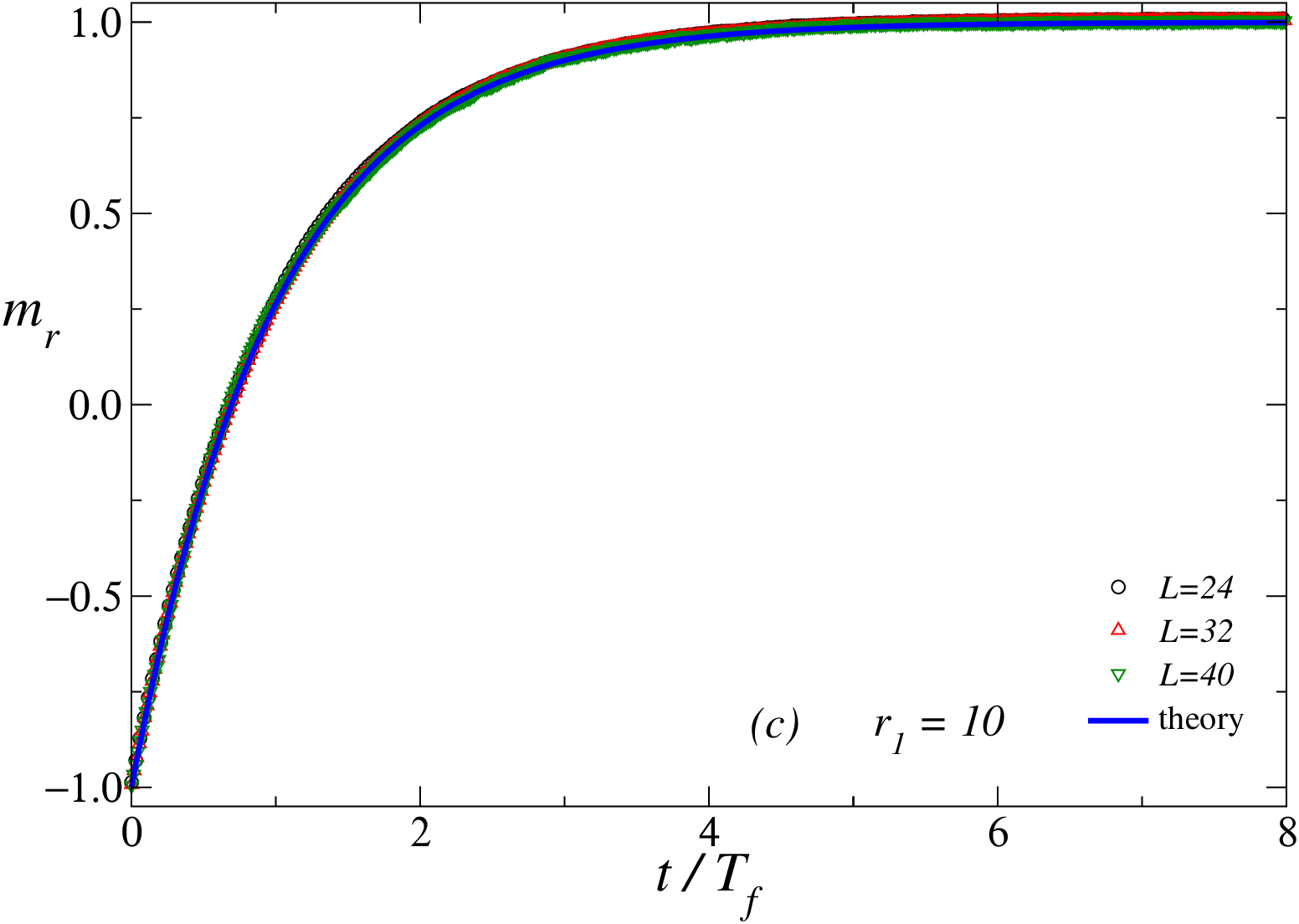} &
\includegraphics*[scale=\graphicscale,angle=0]{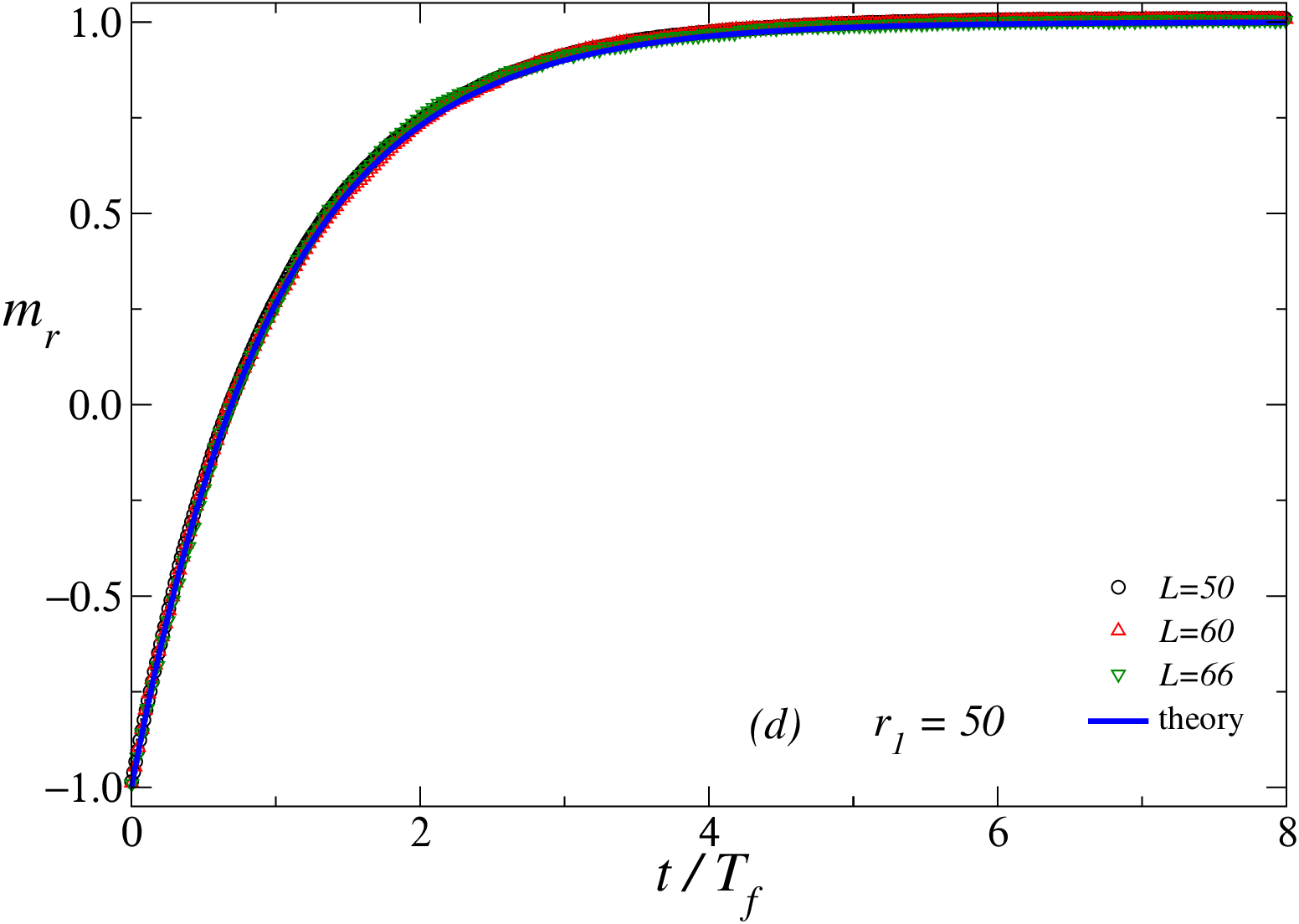} \\
\end{tabular}
%=% \vskip-5mm
\caption{Renormalized magnetization $m_r(t)$ versus 
$t/T_{f}(0.9,r_1,L)$. (a): $r_1 = -1$; (b): $r_1 = 1$;
(c): $r_1 = 10$; (d): $r_1 = 50$. The thick line going
through the points is the theoretical prediction (\ref{mr-th}).
In all cases $T = 0.9T_c$.
}
\label{fig:mt}
\end{figure*}

As a last test of the general theory, we verify prediction
(\ref{mr-th}) for the renormalized magnetization.  As we have verified
the independence of $T_{f}(\mu,r_1,L)$ on $\mu$, we take
$T_{f}(0.9,r_1,L)$ as time scale. In Fig.~\ref{fig:mt} we report
results for $r_1 = -1,1,10,50$ and several values of $L$. We observe
an excellent scaling behavior: data corresponding to different box
sizes fall on top of each other quite precisely. In Fig.~\ref{fig:mt}
we also report the prediction (\ref{mr-th}) (thick lines). It falls on
top of the numerical data, confirming the general coarse-grained
picture of the dynamics.

\subsection{The time scale of the dynamics} \label{sec4.B}

\begin{table}[t]
\begin{tabular}{ccccc}
\hline\hline
$r_1$ & Range   & $\chi^2$/DOF & $a$   & $\alpha$ \\
\hline
4.9   & [20,36] & 3.5/2 & 0.411(7) & 0.1(2) \\
      & [24,36] & 3.4/1 & 0.407(14)& 1.0(4) \\
      & [20,36] & 30/3  &$\sigma$ & 1.61(1) \\
      & [24,36] & 7.3/2 &$\sigma$ & 1.70(2) \\
      & [28,36] & 5.8/1 &$\sigma$ & 1.74(4) \\ \hline
10    & [24,40] & 5.2/2 &0.428(6) & 0.8(2) \\
      & [28,40] & 0.1/1 &0.401(13)& 0.9(4) \\
      & [24,40] & 82/3  &$\sigma$ & 1.56(1) \\
      & [28,40] & 2.8/2 &$\sigma$ & 1.73(2) \\
      & [32,40] & 0.1/1 &$\sigma$ & 1.79(4) \\
\hline\hline
\end{tabular}
\caption{Results of fits to Eq.~(\ref{fitTfp1}) and to
  Eq.~(\ref{fitTfp2}) (here we set $a=\sigma=0.379028...$) for two values
  of $r_1$.  Here $\mu = 0.9$ and $T = 0.9T_c$.  ``Range" gives the
  interval of sizes $L$ considered in the fit, $\chi^2$ is the sum of
  the residuals and DOF is the number of degrees of freedom of the
  fit. }
\label{table-tauL-fits}
\end{table}

In Sec.~\ref{sec3.A} we have performed a careful test of the dynamics
using the first-passage time as the time scale. This allows us to
compare theoretical predictions and numerical data without the need of
tuning any parameter. Here we wish to check the size dependence of
time scale, verifying Eq.~(\ref{tauL}). Using Eqs.~(\ref{TIp}) and
(\ref{tauflsca}), we fit the data of $T_f(\mu=0.9,r_1,L)$ for
$r_1=4.9$ and $r_1=10$ to the ansatz
\begin{equation}
\log T_{f}(\mu=0.9,r_1,L) = a L + \alpha \ln L + b.
\label{fitTfp1}
\end{equation}
If Eq.~(\ref{tauL}) holds, we should find $a = \sigma$, with $\sigma
\approx 0.379028$ for $T = 0.9T_c$, the temperature value of our runs.
The results of the fits are reported in Table~\ref{table-tauL-fits}
(first two rows for each value of $r_1$). We observe that the results,
both for $a$ are and $\alpha$, are significantly size dependent, so it
is difficult to quote a reliable estimate. In any case the estimates
of $a$ apparently approach $\sigma$ as smaller $L$ results are
discarded. Given also the somewhat large statistical error, results
appear to be substantially consistent with the prediction $a =
\sigma$. Then, assuming $a=\sigma$, we can obtain a more precise
estimate of $\alpha$, by fitting the data to
\begin{equation}
\log [T_{f}(\mu,r_1,L) e^{-\sigma L}]  = \alpha \ln L + b.
\label{fitTfp2}
\end{equation}
The estimates of $\alpha$ are reported in Table~\ref{table-tauL-fits}.
Again, we observe a trend with the size $L$ of the systems: as $L$
increases also $\alpha$ increases. Clearly, there are significant
corrections to scaling, as also indicated by the large values of the
$\chi^2$/DOF (DOF is the number of the degrees of freedom of the fit).
It is difficult to quote a final result as data show an increasing
trend with the minimal size used in the fit.  Conservatively, we quote
$\alpha \approx 2$.  Our result is consistent with the result of
Ref.~\cite{BHN-93} that studied the heat-bath dynamics at $h=0$,
finding $\alpha \approx 2.14$ on smaller lattices $L\le 16$.

\subsection{Different dynamics} \label{sec4.C}

Up to now we have only reported results for the checkerboard dynamics.
We wish now to discuss the dynamic behavior observed when using the
sequential and the random dynamics. For this purpose, we have
performed runs with these two different update types at $T = 0.9T_c$,
$r_1 = 10$, and $L=24,28,32$. As before, we analyze the time
dependence of the renormalized magnetization. For both dynamics, we
verify Eq.~(\ref{mr-th}) for the renormalized magnetization,
confirming the universality of the spin-flip dynamics. We can then
compare the efficiency of the different updating procedures, measuring
the ratio
\begin{equation}
   S_{\rm dyn}(r_1,L) = 
  {T_{f,\rm dyn}(0.9,r_1,L)\over T_{f,\rm checker}(0.9,r_1,L) },
\end{equation}
where ``dyn" refers to the sequential and random updates. For the first
type of update we obtain $S_{\rm seq}(r_1, L) =
0.983(5),0.995(5),0.992(6)$, for $L=24,28,32$, respectively, and $r_1
= 10$.  The sequential update is essentially equivalent to the
checkerboard one. For the random update we obtain instead $S_{\rm
  random }(r_1, L) = 4.41(2), 4.48(3),4.53(5)$, for $L=24,28,32$,
respectively.  The random update is clearly slower, but the difference
is only a factor of 4.5.  Note that the time scale of the different
updates differs only by a multiplicative constant. This is at variance
with what happens outside the coexistence region, in which droplets
dominate \cite{RTMS-94} (see also Sec.~\ref{sec5}).

\section{The single-droplet region} \label{sec5}

In the previous sections we have considered the dynamic FSS in the
coexistence region. In that case, one is considering the effective
equilibrium dynamics that consists in flips between the two
essentially degenerate free-energy minima. The relevant phenomenon is
the generation of strip-like domains, while droplet generation does
not play any role.  In this section we consider instead the
intermediate regime in which the phase change can occur either through
strip-like domains or by means of the growth of a droplet. Since the
relevant time scales are proportional to $e^{\sigma L} $ and
$e^{a/h}$, respectively, this regime can be probed by considering the
scaling limit $h\to 0$, $L\to \infty$ at fixed 
\begin{equation}
s=hL.  
\label{sdef}
\end{equation}
In this limit $r_1 \to \infty$ and therefore, for $h>0$, in
equilibrium we have $m\approx +m_0$: if we start from configurations
with $m = -1$, we only observe a single flip to the phase with
positive magnetization. This off-equilibrium dynamics can be described
as discussed in Sec.~\ref{sec3.B}, taking simply $I_- = 0$. All
expressions simplify and we obtain, e.g.,
\begin{equation}
m_r(t) = 1 - 2 \exp \left[ - t/\tau_f(s,L)\right]
\label{mr-sscaling}
\end{equation}
for any value of $s$. While scaling functions are supposed to be independent 
of $s$, any time scale should have a nontrivial $s$ dependence. For instance,
we expect 
\begin{equation}
T_f(\mu,s,L) \sim L^\alpha \exp [A(s) L].
\end{equation}
For the values of $s$ in which the magnetization flip occurs through
the generation of strip-like domains, we should have $A(s) = \sigma$,
while in the regime in which droplets dominate we should find $A(s)
\sim 1/s$. Morever, also $\alpha$ should depend on the regime one is
considering. In particular, Ref.~(\cite{RTMS-94}) predicts $\alpha
\approx 0$ in the single-droplet region for the sequential update and
$\alpha = 1$ for the random update \cite{footnotealpha}.  As before,
we expect the results to be independent of $\mu$ in the scaling limit.

\begin{figure}[tbp]
\begin{tabular}{cc}
\includegraphics*[scale=\graphicscale,angle=-0]{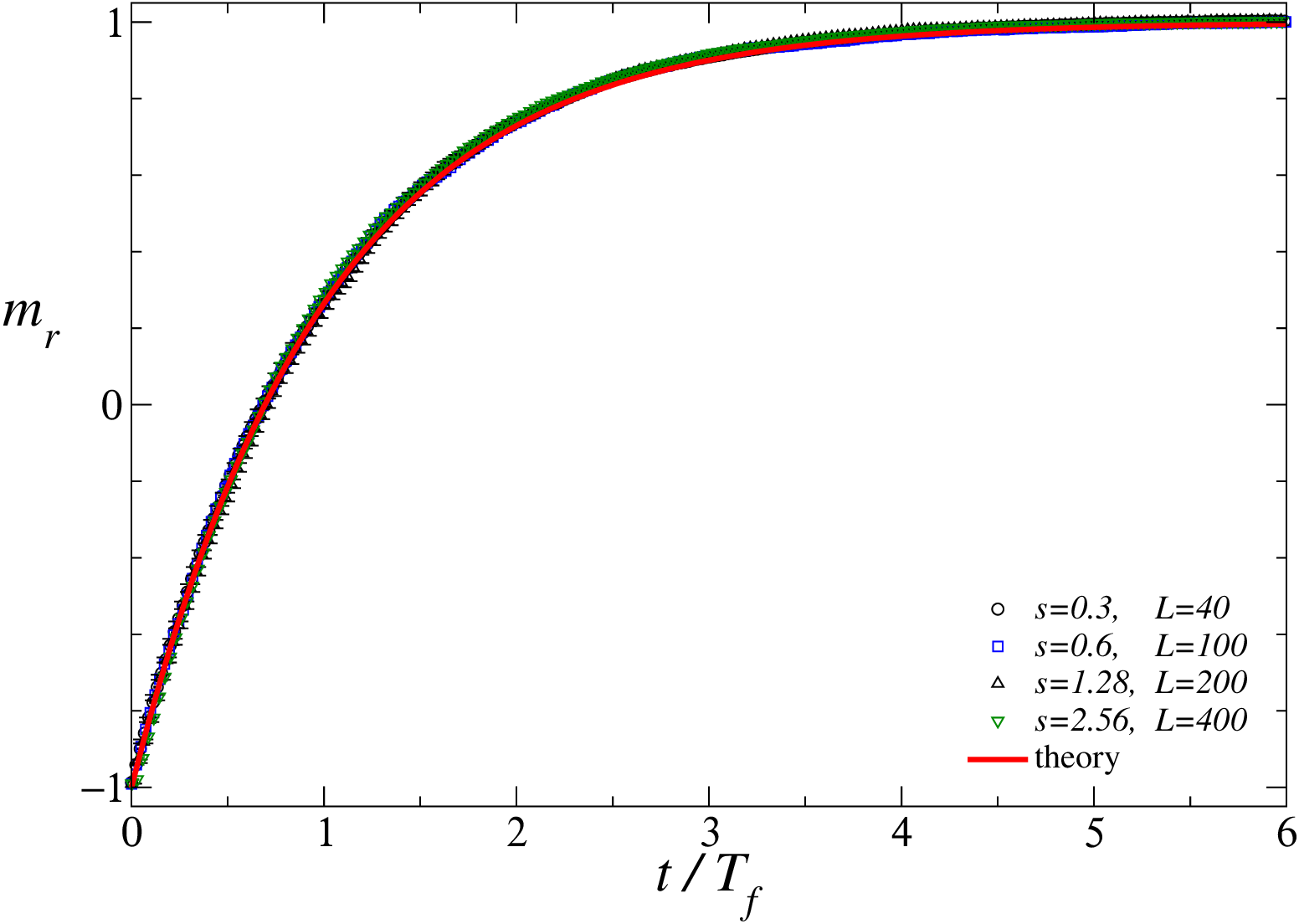} &
\end{tabular}
%=% \vskip-5mm
\caption{Renormalized magnetization $m_r(t)$ versus
  $t/T_{f}(0.9,s,L)$. We report data for different values of $s= hL$; for
  each of them the lattice size $L$ is chosen so that data are in the
  asymptotic scaling regime.  The thick line going through the points
  is the theoretical prediction (\ref{mr-sscaling}).  In all cases $T
  = 0.9T_c$.  }
\label{fig:mr-s-scaling}
\end{figure}

To verify the predicted behavior, we have performed simulations for $s
= 0.3, 0.6,1.28,2.56$ at $T = 0.9T_c$. For each value of $s$ we
consider a few values of $L$ to verify that the size of the system is
large enough to allow us to observe the scaling asymptotic regime.
Results for $m_r(t)$ are reported in Fig.~\ref{fig:mr-s-scaling}. As
expected, all data fall on top of each other and are consistent with
the theoretical prediction (\ref{mr-sscaling}).

We have also studied the behavior of the first-passage time, which
becomes $\mu$ independent as $L$ increases. As before, we use the data
at $\mu = 0.9$ to analyze the $L$ and $s$ dependence of the time
scale. Our data are not precise enough and not sufficiently numerous
to allow us to estimate the exponent $\alpha$. For this reason we have
performed two fits, considering
\begin{equation}
\ln [T_f(0.9,s,L) L^{-\alpha}] = a + A(s) L,
\end{equation}
fixing $\alpha = 0$ (the droplet-region prediction if we assume the
equivalence of the sequential and of the checkerboard update) and 2
(coexistence-region prediction). We obtain $A(s) = 0.238, 0.114,
0.053, 0.025$ for $\alpha = 0$ and $s=0.3$, 0.6, 1.28, and 2.56,
respectively.  For the same values of $s$ and for $\alpha = 2$ we have
$A(s) = 0.161, 0.091, 0.042, 0.019$. Statistical errors are
significantly smaller than $10^{-3}$. Note that all results satisfy
the approximate scaling $A(s) \sim 1/s$, indicating that for these
values of $s$ droplet formation is the relevant mechanism.  It is also
interesting to note that the $\chi^2$ of the fit is significantly
smaller for $\alpha = 0$ than for $\alpha = 2$, in agreement with the
general results of Ref.~\cite{RTMS-94} on the exponent $\alpha$.

\section{The Ising model with a magnetic field on a small lattice domain}
\label{sec6}

It is also interesting to study the dynamics when one considers a
magnetic field that is present only on a small subset of sites.
Specifically, we consider again Hamiltonian (\ref{isiham}), replacing
the magnetic term $h \sum_i s_i$ with $\sum_i h_i s_i$.  We consider
here two cases: (i) the magnetic field is present only on a single
site, that is $h_i$ is always zero except at a single lattice point;
(ii) $h_i$ is nonvanishing only on a lattice line.

\subsection{Magnetic field on a site}

\begin{figure}[tbp]
\begin{tabular}{cc}
\includegraphics*[scale=\graphicscale,angle=-0]{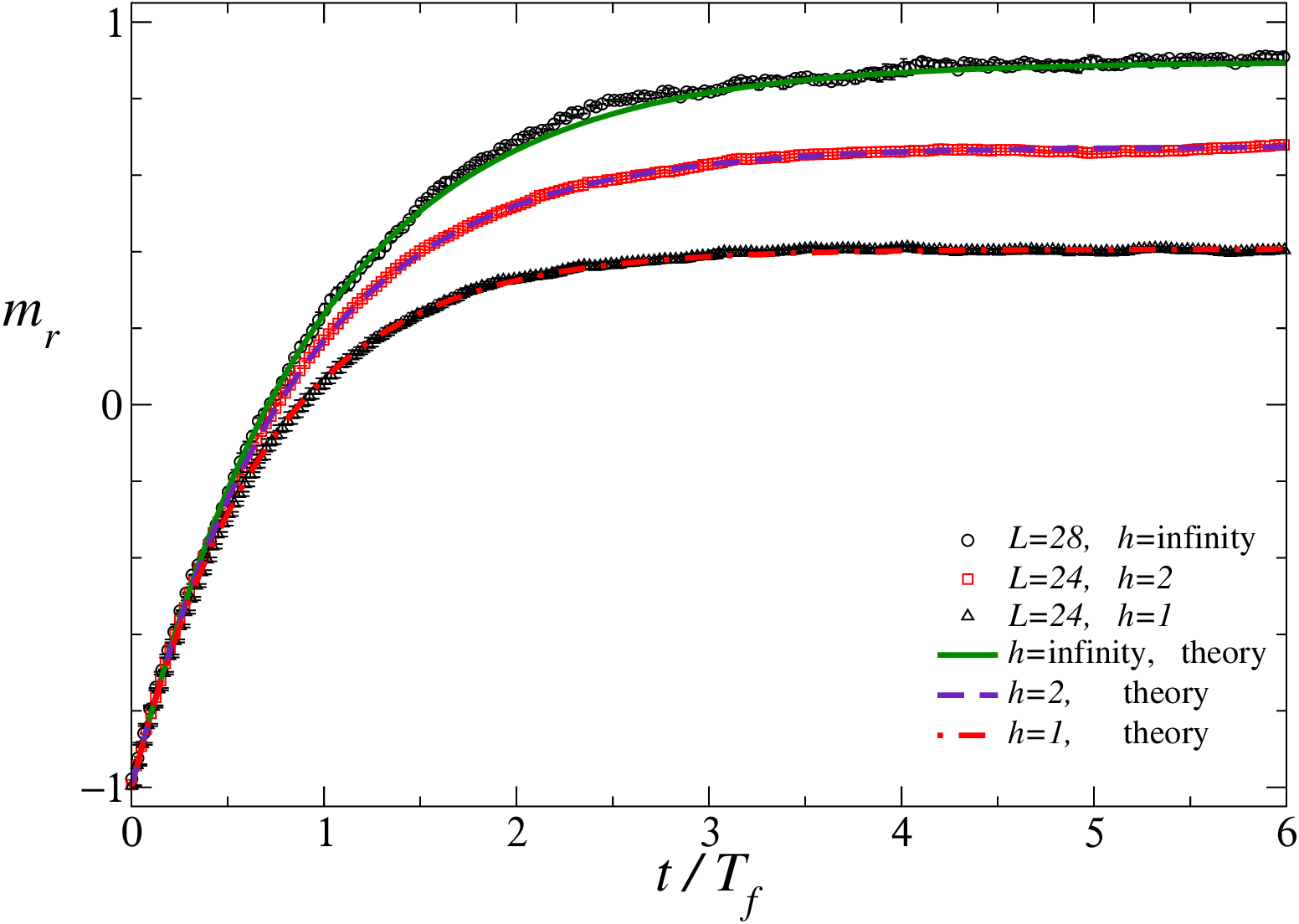} &
\end{tabular}
%=% \vskip-5mm
\caption{Renormalized magnetization $m_r(t)$ versus
  $t/T_{f}(0.9,h,L)$. We report data for different values of the
  magnetic field $h$ on a single site; for each of them the lattice
  size $L$ is chosen so that data are in the asymptotic scaling
  regime.  The thick lines going through the points are the
  theoretical prediction (\ref{mr-th-singlesite}).  In all cases $T =
  0.9T_c$.  }
\label{fig:mr-hsite}
\end{figure}

In the low-temperature phase the addition of a magnetic field on a
single lattice point is enough to break the ${\mathbb Z}_2$ invariance
of the model, thereby generating a finite magnetization. A simple
calculation gives, see App.~\ref{AppA},
\begin{equation}
m_r = \hat{f}_{\rm eq}(h) = m_0 \tanh \beta h.
\label{mr-singlesite}
\end{equation}
Note that for any value of $h$, the absolute value of  $|m_r|$ is always 
less than 1, so that the system is always in the crossover region.
The arguments of Sec.~\ref{sec3.B} should then apply for any $h$. 
Taking into account the different expression for the equilibrium 
magnetization, we obtain 
\begin{eqnarray}
&&m_r(t) = \hat{f}_{\rm eq} (h) - [1 + \hat{f}_{\rm eq} (h)] e^{-t/T_s},
\label{mr-th-singlesite} \\
&& T_s =  {1\over 2} \tau_f(h,L) \left[ 1 + \hat{f}_{\rm eq}(h)\right].
\nonumber 
\end{eqnarray}
In Fig.~\ref{fig:mr-hsite} we show the results for $m_r(t)$ for
$h=1,2,\infty$.  Scaling holds and results are perfectly consistent
with Eq.~(\ref{mr-th-singlesite}). We have also verified that
$T_f(\mu,h,L)$ is independent of $\mu$ and scales as in the case of a
uniform magnetic field at fixed $r_1$. We consider the ratio
\begin{equation}
R(h,L) = {T_f(0.9,h,L)\over T_f(0.9,r_1=1,L)},
\end{equation}
where $T_f(0.9,r_1=1,L)$ is the first-passage time for a uniform magnetic 
field with $r_1 = h L^2 = 1$. 
We obtain for $h = \infty$ $R(h,L) = 0.175(1),0.175(1),0.174(1),0.176(4)$
for $L=16,20,24,28$, respectively. For $h = 1$ we obtain 
analogously $R(h,L) = 0.929(4),0.911(4),0.930(8)$, for 
$L=16,20,24$. The ratio is independent of $L$, indicating that 
the first-passage time scales identically in the two cases.

\subsection{Magnetic field on a line}

\begin{figure}[tbp]
\begin{tabular}{cc}
\includegraphics*[scale=\graphicscale,angle=-0]{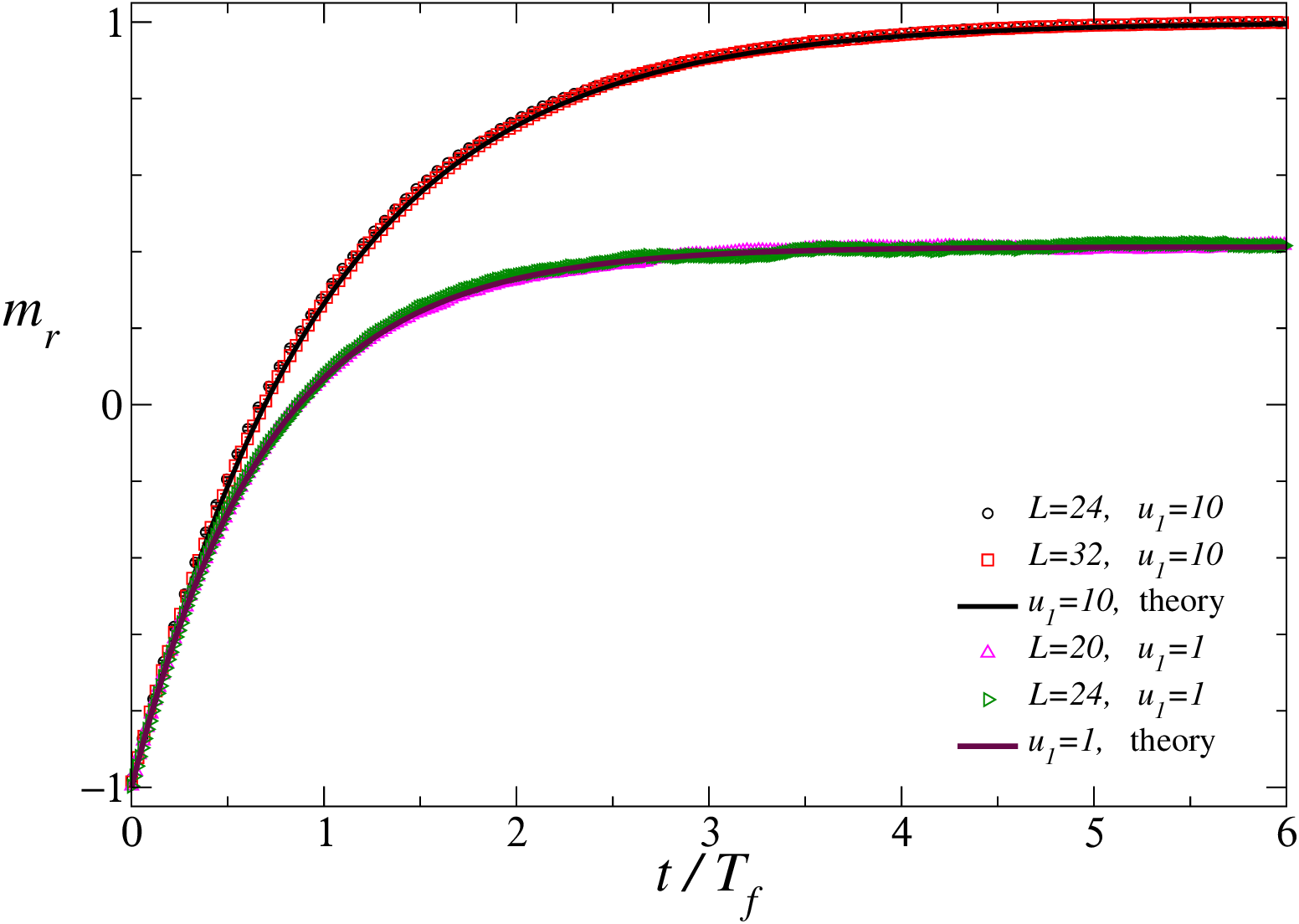} &  
\end{tabular}
%=% \vskip-5mm
\caption{Renormalized magnetization $m_r(t)$ versus
  $t/T_{f}(0.9,u_1,L)$. We report data for different values of $L$ and
  of $u_1 = h L$, where $h$ is nonvanishing only on a lattice line.
  The thick lines going through the points are the theoretical
  predictions.  In all cases $T = 0.9T_c$.  }
\label{fig:mr-hline}
\end{figure}

We now consider the case in which the magnetic field is nonvanishing
only on a lattice line, for instance, on all lattice points $(x,y)$
such that $y=1$. It turns out that the relevant scaling variable is
\begin{equation}
u_1 = h L. 
\label{u1def}
\end{equation}
As $L$ increases, the estimates of $m_r(t)$ at fixed $u_1$ fall onto a
single scaling curve. Moreover, we verify that the equilibrium value
of the magnetization is still given by Eq.~(\ref{feqs}) with $u_1$
replacing $r_1$. The general discussion of Sec.~\ref{sec3.B} applies
also to this case and indeed, the results for $m_r(t)$ are consistent
with Eq.~(\ref{mr-th}) by simply replacing $r_1$ with $u_1$, see
Fig.~\ref{fig:mr-hline}.

We have also investigated the $L$ dependence of the first-passage time,
considering the ratio
\begin{equation}
R(u_1,L) = {T_f(0.9,u_1,L)\over T_f(0.9,r_1,L)},
\end{equation}
between the first passage time for the case of a magnetic field on a
line and that for a uniform field. We always take $u_1 = r_1$, so that
the compared systems have the same equilibrium value of the
magnetization.  For $u_1 = r_1 = 10$, we find $R(u_1,L) = 0.140(1),
0.125(1), 0.113(1)$ for $L = 24,28,32$, respectively, while for $u_1 =
r_1 = 1$ we have $R(u_1,L) = 0.975(6), 0.960(8)$ for $L=20,24$.  The
ratio apparently decreases with $L$. A fit of the data with $u_1 = r_1
= 10$ gives $R(u_1,L)\sim L^{-\beta}$, $\beta = 0.75(10)$.  This
indicates that, at fixed $u_1=r_1$, i.e., for the same equilibrium
value of the magnetization, the dynamics is faster when $h$ is
nonvanishing only on one line than for a uniform magnetic field.

\section{The Potts case: Definitions} \label{sec7}

To test whether the observed behavior in the Ising case is generic,
i.e., it is typical of any FOT, we study a second model that shows a
thermal, i.e., temperature-driven, FOT.  We consider the 2D $q$-state
Potts model on a square lattice.  Its Hamiltonian reads
\begin{equation}
H =  - \sum_{\langle {\bm x}{\bm y}\rangle} \delta(s_{{\bm x}}, s_{ {\bm y}}),
\label{potts}
\end{equation}
where the sum is over the nearest-neighbor sites of a square lattice,
$s_{\bm x}$ ({\em color}) are integer variables $1\le s_{{\bm x}} \le
q$, $\delta(a,b)=1$ if $a=b$ and zero otherwise.  It undergoes a phase
transition~\cite{Baxter-book,Wu-82} at
\begin{equation}
\beta_c = \ln(1+\sqrt{q}), \qquad\qquad T_c = 1/\beta_c,
\end{equation}
between disordered and ordered phases.  The transition is of first
order for $q>4$.  We consider $L\times L$ square lattices with
periodic boundary conditions (PBC), which preserve the $q$-permutation
symmetry.  In infinite volume the energy density $E = \langle H
\rangle/L^2$ is discontinuous at $T_c$, with different $E_c^\pm \equiv
E(T_c^\pm)$.  We define a {\em renormalized} energy density
\begin{eqnarray}
E_r \equiv
\Delta_e^{-1}\,(E -E_c^-),\qquad \Delta_e \equiv E_c^+ -
E_c^-,
\label{ener}
\end{eqnarray}
which satisfies $E_r=0,1$ for $T\to T_c^-$ and $T\to T_c^+$,
respectively. 

Close to the transition, the system shows FSS in terms of the scaling
variable
\begin{equation}
r_1 = L^d \delta, \qquad\qquad  \delta \equiv \beta/\beta_c-1.
\end{equation}
In this limit the finite-size energy density scales as~\cite{PV-17}
\begin{equation}
E_r(T,L) \approx {\cal E}_{\rm eq}(r_1) = (1 + q \,e^{X})^{-1} 
\label{Er-FSS}
\end{equation}
with $X = \Delta_e \beta_c r_1$.

In the following we focus on the case $q=20$, but any other values of
$q>4$ is expected to show analogous behaviors at the FOT. For $q=20$
we have~\cite{Baxter-book} $E(T_c^+) = - 0.626530...$, $E(T_c^-) = -
1.820584...$ We will also be interested in the interface tension,
which takes the value \cite{BNB-93,BJ-92,TB-12} $\beta_c
\kappa=0.185494...$ for $q=20$.

We consider a heat-bath dynamics at fixed $T < T_c$. We use the
checkerboard update and we start the dynamics from a fully disordered
configuration. In the evolution we measure the energy $H(t)$, which
allows us to define the average renormalized energy, using
Eq.~(\ref{ener}) and defining $E = \langle H \rangle/L^2$. As we use
PBC the magnetization
\begin{eqnarray}
M_k = {1\over L^2} \langle \sum_{\bm x} \mu_k({\bm x}) \rangle,
\qquad \mu_k({\bm x}) \equiv {q \delta(s_{\bm x},k) - 1\over q-1},
\label{mkdef}
\end{eqnarray}
vanishes for any value of $T$. To investigate the magnetic properties 
we consider 
\begin{equation}
I_G = L^{-2}  \sum_{k=1}^q  \sum_{{\bm x},{\bm y}}
\langle \mu_k({\bm x}) \mu_k({\bm y}) \rangle.
\label{Igdef}
\end{equation}
In the infinite-volume limit and for $T < T_c$, we have 
\begin{equation}
I_G = {q L^2 m_0^2 \over q-1 }, 
\end{equation} 
where $m_0$ is the spontaneous magnetization, which can be defined by
introducing an infinitesimal breaking of the $q$ state symmetry.
For $q = 20$, we have \cite{Baxter-book,Wu-82} $m_0 = 0.941175...$

\section{The Potts case: scaling arguments} \label{sec8}

The scaling arguments presented for the Ising case extend without
changes to the Potts transition. As before, we define a time scale
\begin{equation}
\tau(L) = L^\alpha \exp(\sigma L)
\label{tau-Potts}
\end{equation}
so that, in the FSS limit, the dynamics in a finite
volume can be parametrized by using $r_1$ and $r_2 = t/\tau(L)$ as
scaling variables.

Also in the Potts case we can perform the coarse graining of the
dynamics.  Indeed, we can assume that the system starts in the
high-$T$ phase and then it suddendly jumps in any of the
equivalent $q$ magnetized states. Therefore, Eq.~(\ref{eqdiff-n})
holds, provided we identify $n(t)$ as the fraction of magnetized
systems at time $t$. Since ${\cal E}_r(t) = 1 - n(t)$, we obtain
\begin{equation}
{\cal E}_r(t) = {I_-\over \lambda} +
    {I_+\over \lambda} e^{-\lambda t},
\label{Ert-Potts-pred}
\end{equation}
with $\lambda = I_+ + I_-$.
For $t\to \infty$ we should recover Eq.~(\ref{Er-FSS}), which implies
\begin{equation}
  {I_-\over I_+} = {1\over q} e^{-\beta_c \Delta_e r_1}
\end{equation}
It follows
\begin{eqnarray}
&& {\cal E}_r(t)= {\cal E}_{\rm eq}(r_1) +
{q  \over q + e^{-\beta_c \Delta_e r_1} }  e^{-t/T_p},\qquad 
\label{EPotts}\\
&& T_p = {q\over I_+ (q + e^{-\beta_c \Delta_e r_1})},
\label{tpdef}
\end{eqnarray}
where ${\cal E}_{\rm eq}(r_1)$ is the static FSS function
(\ref{Er-FSS}).  The quantity $I_G$ can be predicted as well. In the
coarse-grained dynamics the combination
\begin{eqnarray}
&& I_{Gr}(t) = {q-1\over q m_0^2 L^2} I_G(t)
\end{eqnarray}
is equivalent to $1 - {\cal E}_r$, so that 
\begin{eqnarray}
\label{IGPotts}
&& I_{Gr} = {q\over q + e^{-\beta_c \Delta_e r_1} }  
\left( 1 - e^{-t/T_p}\right).
\end{eqnarray}

\section{The Potts case: Monte Carlo results} \label{sec9}

\begin{figure*}[tbp]
\begin{tabular}{cc}
\includegraphics*[scale=\graphicscale,angle=0]{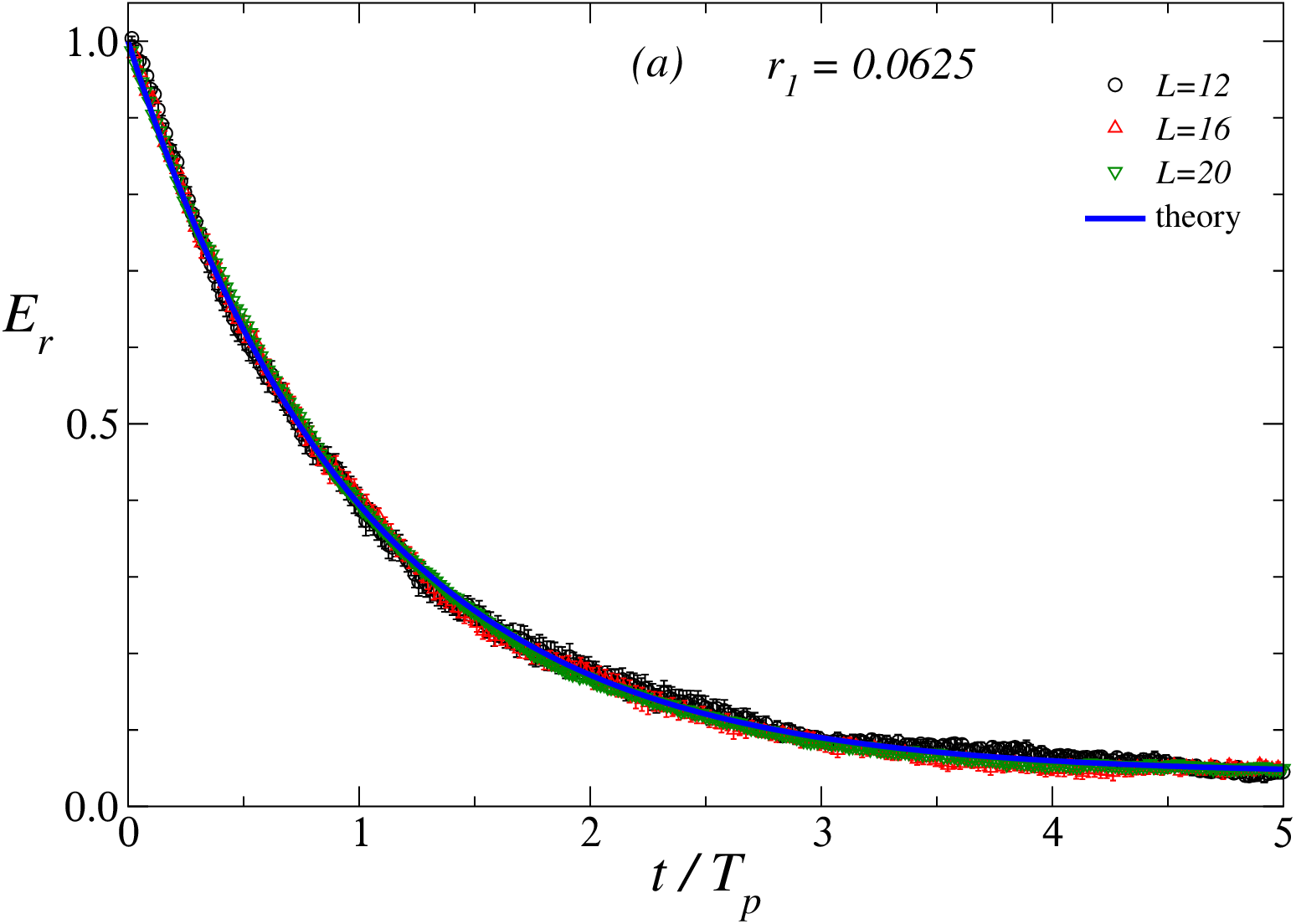} &
\includegraphics*[scale=\graphicscale,angle=0]{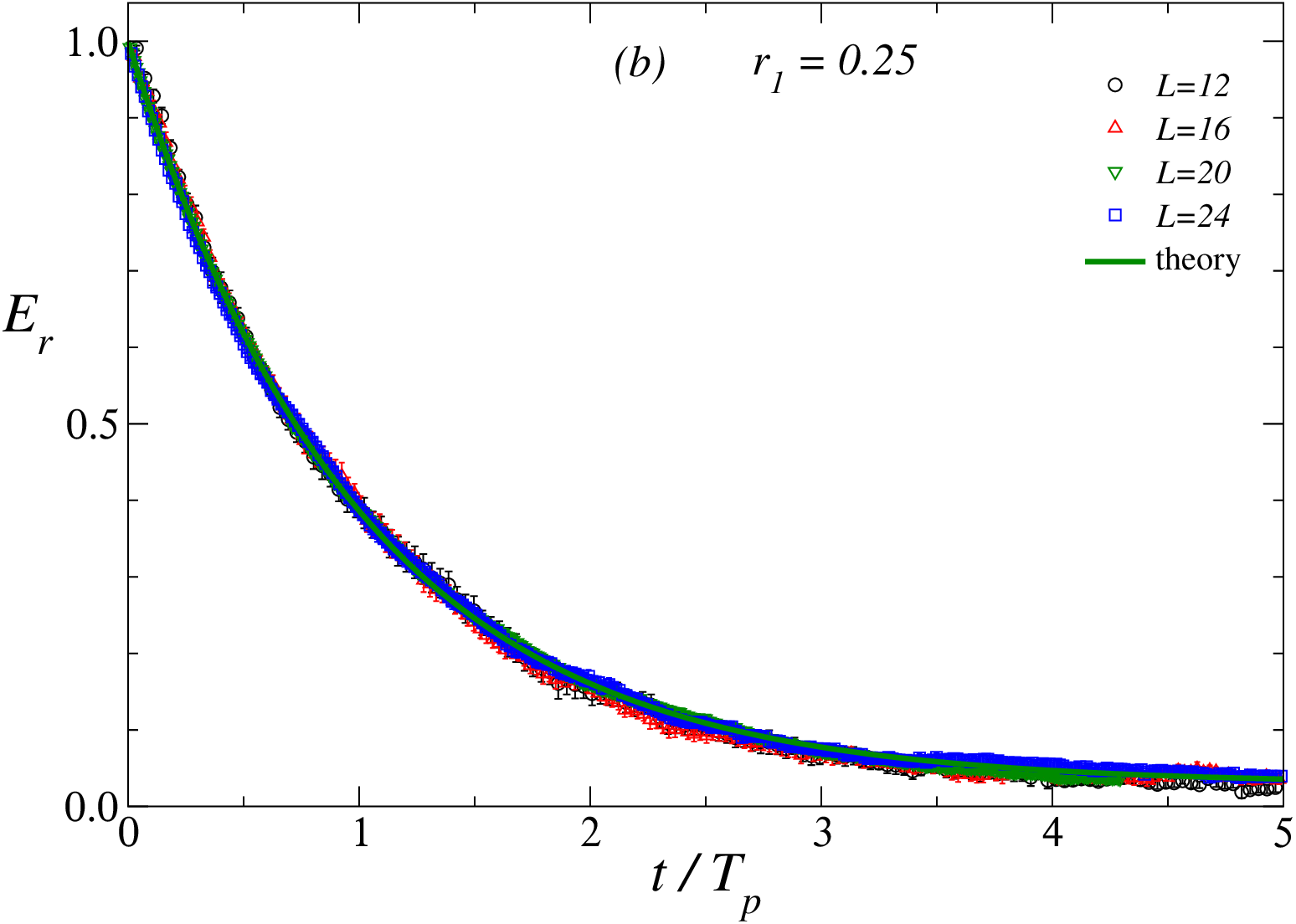} \\
\includegraphics*[scale=\graphicscale,angle=0]{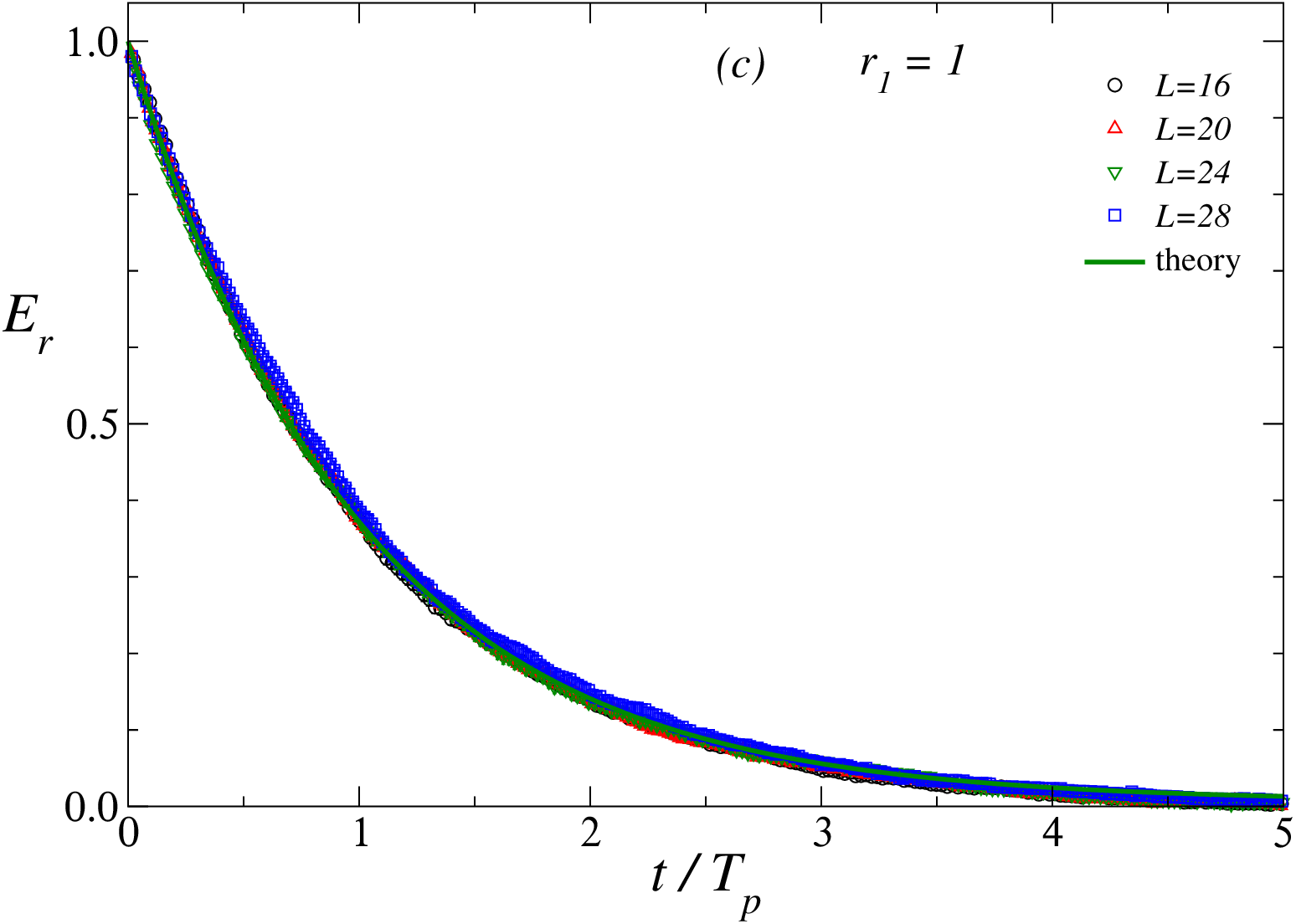} &
\includegraphics*[scale=\graphicscale,angle=0]{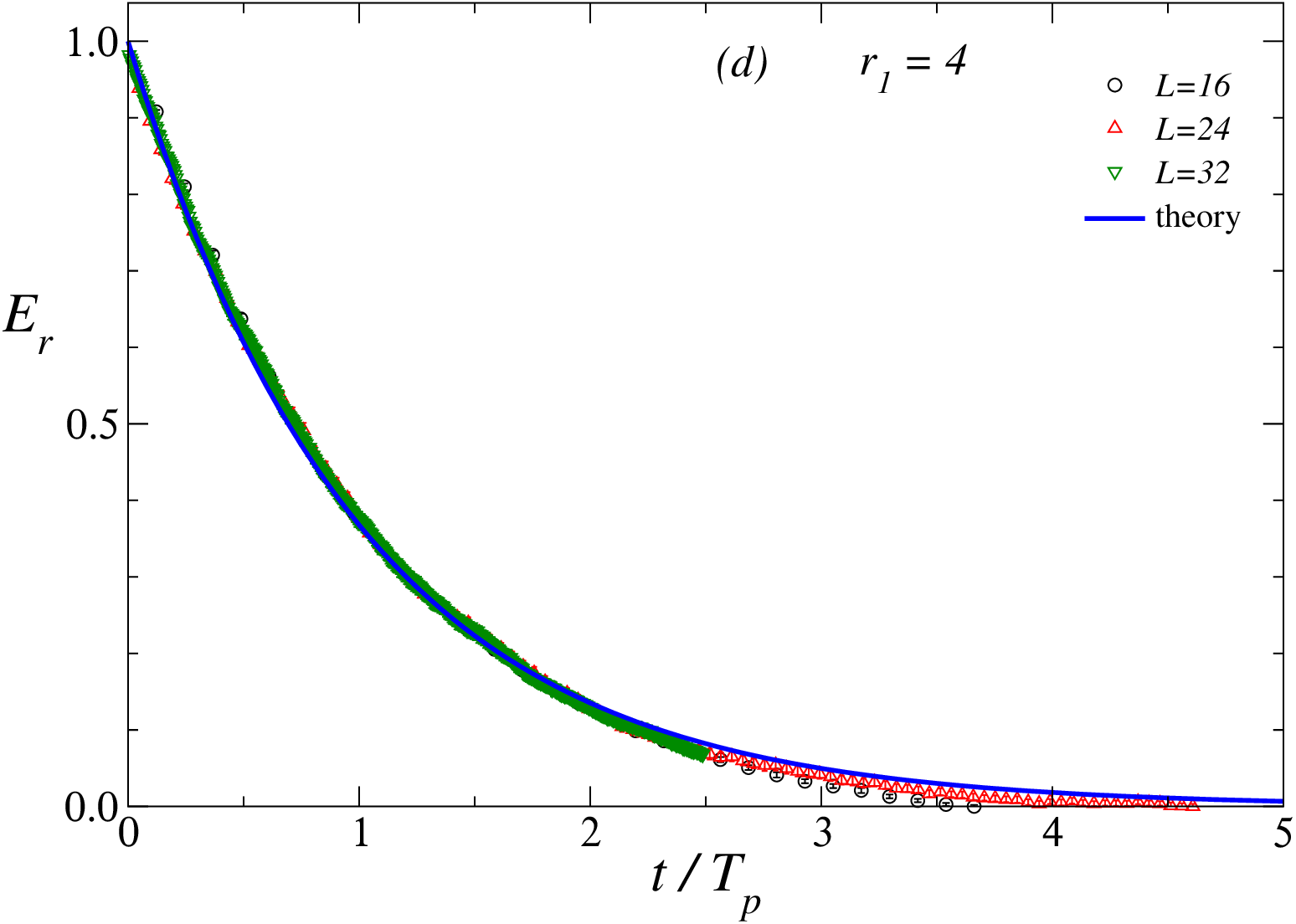} \\
\end{tabular}
\caption{Estimates of ${E}_r$ versus $t/{T}_p$ for 
$r_1 = 0.065$ (a), $r_1=0.25$ (b), $r_1=1$ (c), $r_1=4$
(d), for several values of $L$. The thick line in each panel is the
theoretical prediction (\ref{EPotts}).
}
\label{fig-Potts}
\end{figure*}

To test the general theory, we perform
Monte Carlo simulations for four different values of $r_1$, $r_1 =
1/16, 1/4, 1, 4$, respectively, varying the system size $L$ from 12 to
40. As a first test, we verify Eq.~(\ref{EPotts}), considering
the quantity
\begin{equation}
\widetilde{E}(t) = {1\over q} e^{-\beta_c \Delta_e r_1} \left[
    (1 + q e^{\beta_c \Delta_e r_1}) {\cal E}_r - 1 \right].
\end{equation}
According to Eq.~(\ref{EPotts}), it should behave as a pure
exponential, i.e., $\widetilde{E}(t) = e^{-t/{T}_p}$, in the scaling
limit.  Data accurately satisfy this behavior.

We then fit the data to $\log \widetilde{E}=a\, t$, obtaining
estimates of ${T}_p$.  In Fig.~\ref{fig-Potts} we report the data of
${E}_r(t)$ as a function of $t/{T}_p$ and compare them with the
theoretical prediction (\ref{EPotts}). We observe perfect scaling:
data fall on top of each other for different values of $L$ and are
fully consistent with Eq.~(\ref{EPotts}). Very good scaling is also
observed for $I_G(t)$.  Data behave in full agreement with
Eq.~(\ref{IGPotts}).

\begin{table}[t]
\caption{Results of fits of ${T}_p$ to Eq.~(\ref{fitTfp1}) and to
  Eq.~(\ref{fitTfp2}) (here we set $a=\sigma=0.370988...$) for
  different values of $r_1$.  ``Range" gives the interval of sizes $L$
  considered in the fit. }
\label{table-tauL-fits-Potts}
\begin{tabular}{cccc}
\hline\hline
$r_1$ & range   &  $a$   & $\alpha$ \\
\hline
0.25   & [12,24] &  0.38(2) & 1.4(3) \\
       & [12,24] &  $\sigma$& 1.54(2) \\
1.0    & [16,32] &  0.37(3) & 1.5(6) \\
       & [20,32] &  0.32(6) & 2.7(1.4) \\
       & [16,32] &  $\sigma$& 1.56(4) \\ 
       & [20,32] &  $\sigma$& 1.58(6) \\ 
       & [24,32] &  $\sigma$& 1.34(13) \\ 
4.0    & [16,40] &  0.41(2) & 0.0(5) \\
       & [20,40] &  0.39(3) & 0.6(8) \\
       & [16,40] &  $\sigma$& 0.96(9) \\ 
       & [20,40] &  $\sigma$& 1.07(9) \\ 
       & [24,40] &  $\sigma$& 1.16(14) \\ 
       & [28,40] &  $\sigma$& 1.26(18) \\ 
\hline\hline
\end{tabular}
\end{table}

Finally, we verify the size dependence of the scale, performing the
same fits as we did in the Ising case. We consider the time scale
${T}_p$ and first perform fits to Eq.~(\ref{fitTfp1}). Results are
reported in Table~\ref{table-tauL-fits-Potts}. For all values of $r_1$
the constant $a$ is consistent with $\sigma = 2 \beta \kappa \approx
0.371$, confirming the theoretical prediction (\ref{tau-Potts}). To
estimate $\alpha$, we perform fits to Eq.~(\ref{fitTfp2}), using the
theoretical prediction for $\sigma$. For $r_1 \approx 0.25$ and 1,
results give $\alpha \approx 1.5$. Results for $r_1 = 4$ are also
consistent: the estimates of $\alpha$ are lower, but show a
significant increasing trend.

\section{Potts model: scaling in the presence of a single strongly
ferromagnetic bond} \label{sec10}

\begin{figure}[tbp]
\begin{tabular}{cc}
\includegraphics*[scale=\graphicscale,angle=-0]{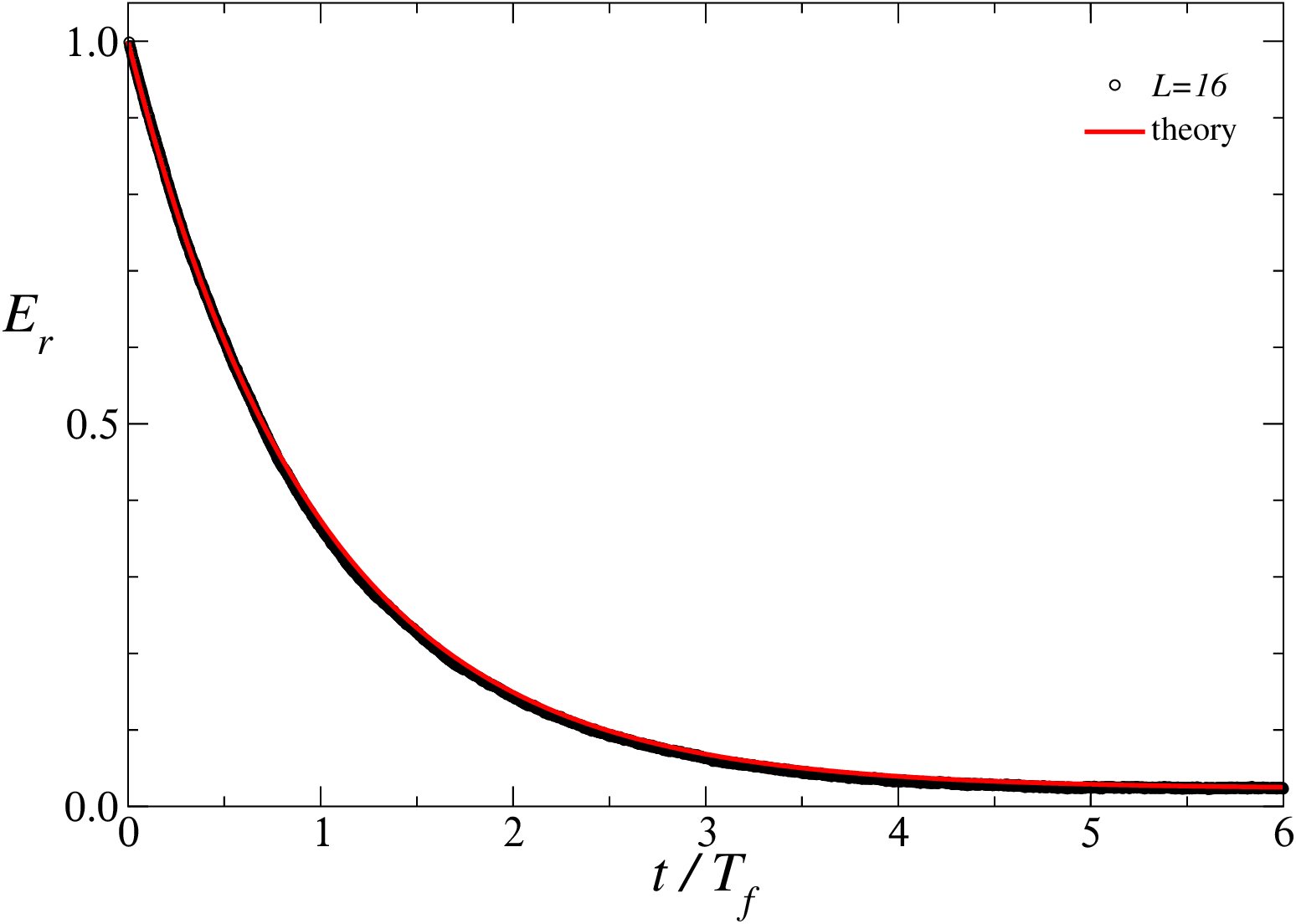} &  
\end{tabular}
%=% \vskip-5mm
\caption{Renormalized energy $E_r(t)$ versus
  $t/T_{f}(0.1,L)$, where $T_f(0.1,L)$ is the first-passage time
  corresponding to $E_r = 0.1$. Simulation with $\beta' = 2\beta$ and
  $L=16$.  The thick line going through the points is the theoretical
  prediction.  }
\label{fig:Er-Potts-link}
\end{figure}

The analysis we have presented in Sec.~\ref{sec6} for the Ising model
with a magnetic field different from zero only on a subset of lattice
points can be extended to the Potts model. For instance, one can
consider a Hamiltonian with a single strongly ferromagnetic bond,
i.e., such that
\begin{equation}
\beta {\cal H} = - \beta \sum_{\langle {\bm x}{\bm y}\rangle} 
   \delta(s_{{\bm x}}, s_{ {\bm y}}) - (\beta'-\beta) 
     \delta(s_{{\bm a}}, s_{ {\bm b}}),
\end{equation}
where $s_{{\bm a}}$ and $s_{{\bm b}}$ are the spins at the vertices of
a lattice bond $\langle {\bm a}{\bm b}\rangle$. Here we set $\beta =
\beta_c$ as we wish to investigate the behavior at the thermal FOT and
consider $\beta'\not=\beta_c$. The equilibrium value of the energy in
the FSS limit can be computed exactly, see App.~\ref{AppB}. Note that
$0< E_r < 1$ for any value of $\beta'$ (even for a negative value),
indicating that the system is always in the coexistence region.  We
can therefore apply the arguments of Sec.~\ref{sec8}. In
Fig.~\ref{fig:Er-Potts-link} we report numerical data for $\beta' = 2
\beta_c$. Results scale as predicted by Eq.~(\ref{Ert-Potts-pred}),
provided we fix the ratio $I_-/I_+$ using the equilibrium value
(\ref{Er-eq-singlebond}).

\section{Conclusions} \label{sec11}

We have investigated the dynamic behavior of finite-size systems close
to a FOT. Static quantities obey general FSS laws when expressed in
terms of the scaling variable $r_1 = \delta L^d$, where $d$ is the
space dimension and $\delta$ specifies the distance from the
transition point. At magnetic transitions we set $\delta = h$, where
$h$ is the magnetic field, while at thermal transitions one can take
$\delta = \beta/\beta_c - 1$. If one considers the limit $\delta \to
0$ and $L\to \infty$ at fixed $r_1$, one is always probing the
coexistence region. Therefore, for periodic boundary conditions, or
more generally for boundary conditions that do not favor a specific
phase, the system oscillates among the different coexisting phases as
the corresponding free energy barrier is finite. The relevant time
scale $\tau(L)$ is the tunnelling time between the coexisting phases,
which scales as $\tau(L) \sim L^\alpha \exp(\sigma L)$, where $\sigma$
is proportional to the interface tension, and $\alpha$ is an
appropriate exponent.

We develop a DFSS theory for the dynamic behavior in this regime,
characterized by the coexistence of the two phases.  If we consider
time scales of the order of $\tau(L)$, the dynamic behavior can be
described by using a two-state coarse-grained (Poisson) dynamics.
This allows us to obtain exact predictions for the dynamical scaling
functions.

The arguments that we present are general and therefore they should
apply to any FOT with a discrete order parameter. Systems with
continuous order parameters are expected to behave differently,
because of the presence of Goldstone modes (see, e.g.,
Ref.~\cite{PV-16}, for a discussion).

We test these ideas in the 2D Ising and $q$-state Potts models. In the
first case, we consider the magnetic FOTs that occur in the
low-temperature phase for $h = 0$. We consider a purely relaxational
dynamics at fixed $h$ and $T$, starting from a completely ordered
configuration.  We investigate the behavior for a uniform magnetic
field and for a magnetic field that vanishes everywhere except on a
lattice point or a lattice line. In the Potts case we set $q= 20$ and
we consider the thermal FOT that is observed by varying the
temperature.  In particular, we consider the relaxational evolution
using a heat-bath dynamics at fixed $T<T_c$, starting from a
metastable disordered configuration.  The numerical analyses for both
models fully confirm the general picture.

Our study should be of particular relevance for experiments of
moderately small systems (such as those we have considered for our
tests), when the longest time scale of the system is of the order of
the time scale of the experiment, as it may be the case in several
physical contexts.

\appendix 

\section{Ising model: Magnetization in the presence of a magnetic field on a
single site} \label{AppA}

We compute here the magnetization for a finite system in which the
magnetic field is non vanishing only at a single point (for
definiteness we assume a finite $h$ in the origin).  If ${\cal H}_0 $
is the Hamiltonian in the absence of a magnetic field and $\langle
\cdot \rangle_0$ is the average with respect to ${\cal H}_0 $, we
rewrite
\begin{equation}
\langle A \rangle_h  = {\langle A e^{\beta h s_0} \rangle_0 \over 
        \langle e^{\beta h s_0} \rangle_0 },
\end{equation}
where $A$ is an arbitrary function of the spins.
Then, we use the identity 
\begin{equation}
   e^{\beta h s_0} = \cosh \beta h + s_0 \sinh \beta h,
\end{equation}
which follows from the fact that $s_0$ takes only the values $\pm1$.
In the absence of a magnetic field $\langle s_0 \rangle_0$ vanishes and 
therefore we obtain
\begin{equation}
\langle A \rangle_h = \langle A \rangle_0 + 
\langle A s_0\rangle_0 \tanh \beta h,
\end{equation}
for any operator $A$.
For the magnetization it follows 
\begin{equation}
m(h) =  \left\langle s_0 \left({1\over V} \sum_i s_i \right) \right\rangle_0 
\tanh \beta h.
\end{equation}
Using translation invariance we can rewrite 
\begin{equation}
m(h) =  \left\langle \left({1\over V}\sum_i s_i\right )^2 \right\rangle_0 
\tanh \beta h.
\end{equation}
In the absence of a magnetic field the average value is equal to 
$m_0^2$ and therefore 
\begin{equation}
m(h) =  m_0^2 \tanh \beta h.
\end{equation}

\section{Potts model: energy in the presence of an additional 
single-site bond energy term} \label{AppB}

In analogy with the Ising case we now compute the energy for a Potts model
in which there is an additional energy term associated with a single bond. 
More precisely, if $H_0$ is the Potts Hamiltonian (\ref{potts}), we consider 
\begin{equation}
{\cal H} = H_0 - a \delta(s_a,s_b) ,
\end{equation}
where $s_a$ and $s_b$ are the colors at the vertices of an arbitrary 
lattice bond. If $\Delta \beta = \beta a$, and $\langle \cdot \rangle$ and 
$\langle \cdot \rangle_0$ are the averages with respect to Hamiltonians
$\cal H$ and $H_0$, respectively, we have 
\begin{equation}
\langle H_0 \rangle = {\langle H_0 e^{\Delta \beta \delta(s_a,s_b)} \rangle_0
    \over \langle e^{\Delta \beta \delta(s_a,s_b)} \rangle_0 }\; .
\end{equation}
Now, since $\delta(s_a,s_b)$ takes only two values, 0 and 1, we can write 
\begin{equation}
e^{\Delta \beta \delta(s_a,s_b)} = 1 + 2 f \delta(s_a,s_b) \qquad 
f = {1\over 2} (e^{\Delta \beta } - 1).
\end{equation}
Using also the translational invariance of the model with Hamiltonian 
$H_0$ (we assume periodic boundary conditions), it follows 
\begin{equation}
\langle H_0 \rangle ={\langle H_0 \rangle_0 - f \langle H_0^2 \rangle_0/L^2
     \over 1 - f \langle H_0 \rangle_0/ L^2}. 
\end{equation}
If $E = \langle H_0 \rangle/L^2$ is the energy density for $a = 0$, we use 
the identity
\begin{equation}
\langle H_0^2 \rangle_0 = L^4 E^2 - L^2 {\partial E\over \partial \beta},
\end{equation}
and Eq.~(\ref{Er-FSS}) to derive at the critical point 
\begin{equation}
{1\over L^4} \langle H_0^2 \rangle_0 =  {(E_c^+)^2 +  q (E_c^-)^2\over 1 + q}.
\end{equation}
If we define 
\begin{equation}
E_r = \Delta^{-1}_e \left( \langle H_0 \rangle/L^2 - E_c^-\right),
\end{equation}
we obtain
\begin{equation}
E_r = {1 - E_c^+ f\over 1 - E_c^+ f + q (1 - E_c^- f)}.
\label{Er-eq-singlebond}
\end{equation}
Note that $f\ge -1/2$ ($f=-1/2$ is obtained for $a\to -\infty$) 
and $E_c^\pm > -2$, so that $E_r$ satisfies the strict inequality 
$0 < E_r < 1$.


\begin{thebibliography}{99}
\bibitem{footnotealpha} In the notations of Ref.~\cite{RTMS-94}, 
the nucleation time is proportional to $L^{-2} |h|^{-(b+c)} 
\exp (a/h)$. At fixed $hL$ this allows us to predict 
$\alpha = b + c - 2 \approx 0$, if we use the estimate $b + c \approx 2$
valid for the sequential update. For the random update, one expects 
$b + c = 3$ and correspondingly $\alpha = 1$.
\end{thebibliography}
\end{document}